\documentclass[twocolumn,astrosymb,times]{aastex631}
\usepackage{rotating}
\usepackage{booktabs}
\usepackage{textcomp, gensymb}
\usepackage{tabularx}
\usepackage{amsmath}
\usepackage{appendix}

\shorttitle{Spotted Non-accreting Pre-Main-Sequence Stars}

\begin{document}

\title{Spectral Biases, Starspot Morphology, and Dynamo Transitions on the Pre-Main Sequence: Insights from the X-Shooter WTTS Library}

\correspondingauthor{Facundo P\'erez Paolino}
\email{fperezpa@caltech.edu}

\author[0000-0002-4128-7867]{Facundo P\'erez Paolino}
\affiliation{Department of Astronomy, California Institute of Technology, 1216 East California Blvd, Pasadena, CA 91125, USA}

\author[0000-0001-8642-5867]{Jeffrey S. Bary}
\affiliation{Colgate University, 13 Oak Drive, Hamilton, NY 13346, USA}

\author{Lynne A. Hillenbrand}
\affiliation{Department of Astronomy, California Institute of Technology, 1216 East California Blvd, Pasadena, CA 91125, USA}

\author{Benjamin Horner}
\affiliation{Colgate University, 13 Oak Drive, Hamilton, NY 13346, USA}

\author[0000-0002-9540-853X]{Adolfo Carvalho}
\affiliation{Department of Astronomy, California Institute of Technology, 1216 East California Blvd, Pasadena, CA 91125, USA}

\begin{abstract}

{Starspots are ubiquitous in young, low-mass stars, yet their impact on the spectral classification and fundamental parameter inference of pre-main sequence stars (PMS) has been largely overlooked. In this study, we demonstrate that cool starspots systematically distort spectral morphology and bias the effective temperatures, surface gravities, and luminosities derived for non-accreting Weak-Lined T Tauri Stars (WTTS). Using a sample of 56 WTTS with high-resolution, broad-band X-Shooter spectra, we perform two-temperature spectral fits that explicitly account for spot coverages and temperature contrasts. These composite models consistently outperform traditional single-temperature fits, particularly in the 3350–4000~K regime, where spot contributions dominate the red-optical and near-infrared flux. Moreover, we propose that surface gravity discrepancies between optical and infrared measurements are a natural consequence of spot-dominated emission in PMS stars. We find that single-temperature models can overestimate effective temperatures by up to 700~K and underestimate $\rm \log g$ by 1–2 dex. Using spot-corrected effective temperatures, we derive masses and ages from traditional, magnetic, and spotted evolutionary models, finding that spot-corrections systematically raise inferred masses by up to 80\% and stellar ages by up to 0.5 dex. These discrepancies are strongest for stars in the $\rm 0.3–0.8\ M_{\odot}$ range. Using starspots as a proxy for magnetic topology, we find evidence that a shift from largely axisymmetric to non-axisymmetric magnetic fields dominated by small-scale structures coincides with the formation of a radiative core during PMS evolution, effectively distinguishing between the``` convective and interface dynamo regimes.
}

\end{abstract}


\section{Introduction} \label{sec:intro}

Our ability to understand the formation of Sun-like stars and the evolution of their protoplanetary disks into planetary systems requires us to successfully disentangle the complex physical processes associated with the earliest stages of evolution. At the heart of a natal circumstellar disk resides a puffy, still contracting, {pre-main-sequence} (PMS) star \citep{Hartmann2016}. Observations of such systems are complicated by their unresolved nature, whereby spectroscopy and photometry reflect the composite emission of all the different emission components \citep[e.g.,][]{Basri1990, Hartigan1995}. The youngest and least evolved PMS stars are often still feeding from their circumstellar disks via accretion columns culminating in shocks at the stellar surfaces producing excess UV continuum emission \citep{Gullbring1998}. Strong magnetic fields generating outflows, bipolar jets, slower moving wide-angle disk winds, and X-ray flares are ubiquitous. 

A common method for isolating the central emission of the stars relies on the use of theoretical or empirical spectral templates to represent the underlying star \citep{Hartigan1991}. Once an inference of the stellar spectral type is made, often based on line strength ratios, measurements of {“veiling between”} the stellar spectrum and the appropriate template can be made, and the stellar emission can be distinguished from other sources of emission \citep{Fischer2011}. 

{The spectra of non-accreting, weak-line T Tauri Stars {WTTS} have commonly been used as spectral templates to represent the underlying photospheric spectra of their accreting counterparts. WTTS are used because they provide a better match in surface gravity when compared than main sequence stars, while also lacking the systematic issues inherent to stellar models of low mass stars with poorly constrained processes of radiative transfer. They also account for the effects of strong magnetic fields, which most atmospheric models and main sequence stars lack \citep{Valenti1993}.}

PMS stars have strong 1-3 kG magnetic fields \citep[e.g.][]{lopez2021, Flores2022} that can suppress convection near the surface and lead to the formation of large, cool starspots \citep{Strassmeier2009}. Since these regions emit at lower temperatures than the surrounding photosphere, they introduce a second component to the unresolved stellar spectrum that will bias colors towards redder values \citep{Gullbring1998}, generate rotational modulation of lightcurves \citep{Vrba1986}, and lead to spectral type mismatches due in part to the stronger molecular absorption features associated with the spots \citep[e.g.,][]{Vacca2011, Debes2013, Bary2014}. 

Recent efforts by \citet{Gangi} and \citet{PerezPaolino2024} find systematically high spot filling factors (up to $\approx80\%$) for stars in Taurus-Auriga, suggesting that starspot coverage may decrease with age, as is also suggested by a previous analysis of archival photometry \citep{Morris2020}. These models were extended by \citet{PérezPaolino2025} to account for accretion and disk excesses as well as starspot emission in an effort to understand the spectra of the 16 accreting Classical T~Tauri Stars (CTTS) in \citet{Fischer2011}. The results of this study demonstrate that when spots are not accounted for in the IR spectral templates of PMS stars, accretion excesses based on them can be biased by as much as a factor of a few to ten.

Extensive libraries of non-accreting PMS stars covering a large range of spectral types at high resolution and over large wavelength ranges are scarce yet vitally important. The most notable example of such a library is the set of \citet{Manara2013} and \citet{Manara2017} X-Shooter NUV-NIR spectra of non-accreting PMS stars, covering spectral types G5 to M9.5. In its latest iteration, the X-Shooter template library has been improved to also allow for empirical interpolation between spectral templates \citep{Claes2024} and a future version will include accretion shock models capable of simulating spectra of accreting stars \citep[e.g.,][]{Calvet1998, Pittman2022}. 

\citet{Fang2021} noticed spectral type mismatches as large as a few spectral subclasses between the \citet{Manara2013} and \citet{Manara2017} spectral templates and those derived with their spectral typing method. These issues are similar to those found by \citet{Vacca2011, Pecaut2016, Gully2017, PerezPaolino2024} extending to stellar masses and ages. Not only are mass and age discrepancies seen when using HR diagrams to estimate stellar parameters from optical- or NIR-derived temperatures, but the issue persists when using Kiel diagrams with surface gravity estimates from high-resolution optical and NIR spectroscopy. \citet{Flores2019} find systematic differences between surface gravity measurements ($\rm \log g$) for BP~Tau and V437~Aur performed using optical versus infrared spectra, wherein they find infrared surface gravities are higher by up to one dex when compared to optical measurements. They also find systematic differences between infrared and optical temperatures, resulting in spectral type mismatches similar to those mentioned above. While the spectral type mismatches can be explained by the presence of large cool starspots, could they also be responsible for the incongruent surface gravities?

It is becoming clear that, if unaccounted for, the presence of cool spots on the surfaces of PMS stars with large filling factors may significantly impact our ability to accurately infer their ages and masses, determine their surface gravities, and constrain their mass accretion rates. In this study, we fit a grid of synthetic spectra of spotted stars to the non-accreting PMS spectral templates presented in \citet{Claes2024} to search for evidence of and to characterize the nature of spots on the stars in this library. We measure starspot filling factors and temperatures for both the spotted and unspotted photospheric regions of 56 PMS stars in Taurus, Lupus, Upper~Sco, Sigma Orionis, TW~Hya, and Eta Chamaeleonitis. This paper aims to provide a rigorous demonstration of the limitations of current single-temperature classification approaches for PMS stars, and establish the need for a revised framework that incorporates multi-temperature surface structures.


\section{X-Shooter Spectra of PMS Stars}

\citet{Claes2024} presents X-Shooter \citep{Vernet2011} spectra of 56 PMS stars with negligible extinction taken at ESO's Very Large Telescope (VLT) to be used as non-accreting templates in future studies. This sample is a composite of new observations with archival spectra from \citet{Manara2013} and \citet{Manara2017}\footnote{For a description of the observations and data reduction process, we direct the readers to the aforementioned papers.}. All of these sources are believed to be young ($\leq$~10~Myr) and solitary class III sources with no detectable K-band excesses \citep{Lovell2021}.  In Table~\ref{tab:singlefits}, we list each of the sources with spectral types and effective temperatures (Columns 1-4) from \citet{Claes2024}.

\section{Fitting Spectra of PMS Stars}
\begin{figure*}[ht!]
\centering
\includegraphics[width=1\linewidth]{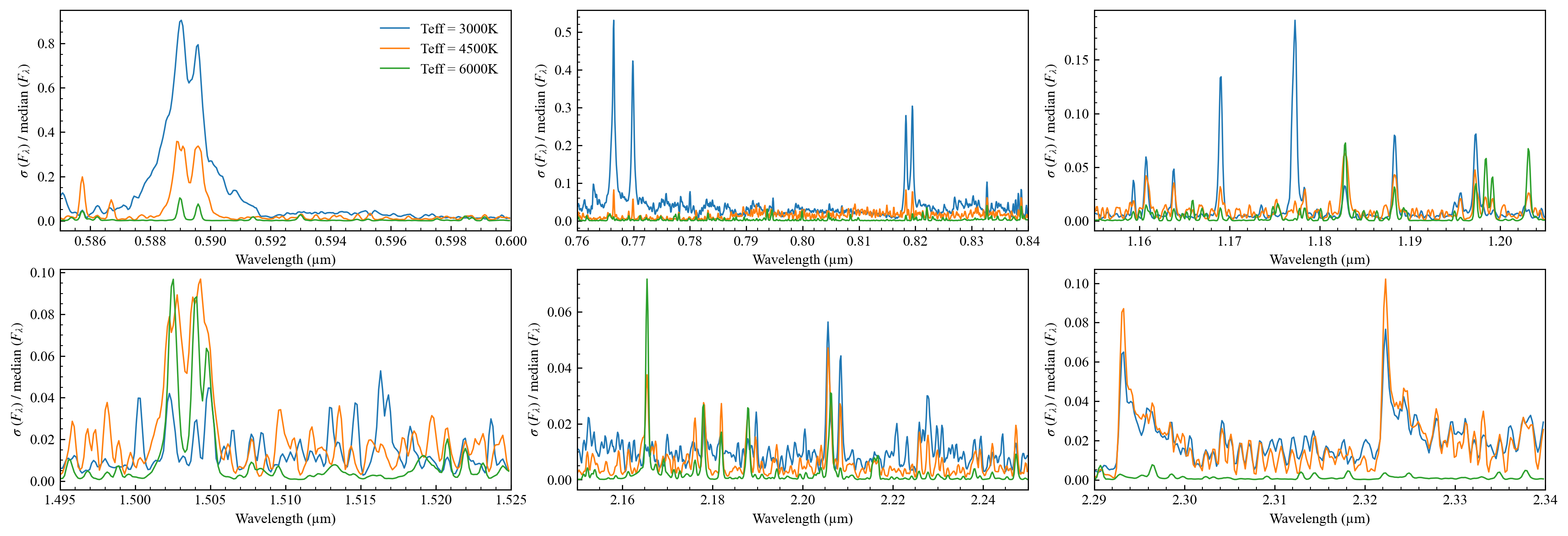}
\caption{\textbf{Surface gravity dispersion spectra for the regions used in fitting.} These are shown for representative temperatures of 3000~K (blue), 4500~K (orange), and 6000~K (green). Discrepant strength variations are seen for different temperatures, indicating the different efficacy of a given region as a surface gravity indicator at that temperature. The spectra were computed at instrumental resolution and broadened by 50 km/s for display purposes and to match typical PMS rotational velocities.}\label{fig:disp_spectra_logg}
\end{figure*}

\begin{figure*}[ht!]
\centering
\includegraphics[width=1\linewidth]{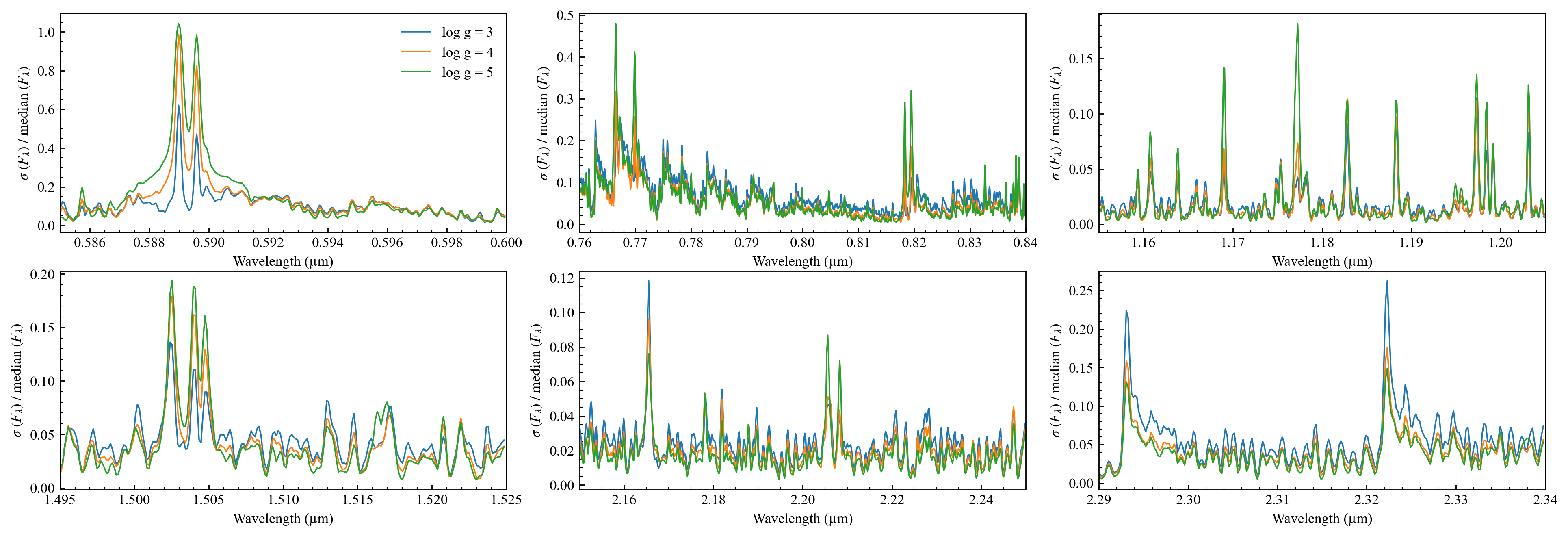}
\caption{\textbf{Temperature dispersion spectra for the regions used in fitting.} Same as Figure~\ref{fig:disp_spectra_logg} but for log~g = 3 (blue), log~g = 4 (orange), and log~g = 5 (green).}\label{fig:disp_spectra_temp}
\end{figure*}

\subsection{Single Temperature Models}\label{sec:singlefits}
X-Shooter spectral templates of PMS stars were fit using BTSettl \citep{allard2014} theoretical atmospheres over discrete wavelength ranges containing gravity and temperature sensitive lines \citep[e.g.,][]{Mohanty2004}. These regions were found by computing dispersion spectra as the standard deviation at each wavelength computed across a set of continuum-subtracted theoretical atmospheres spanning the range $\rm \log g$~=~2.5-5.5 and $T_{\rm eff}=2000-7000~K$, with the continuum subtraction performed using an asymmetric least-squares algorithm \citep{Carvalho2023}. These dispersion spectra represent the fractional change in flux with respect to the continuum as a function of wavelength. {We calculated two sets of dispersion spectra. One in which we fixed $T_{\rm eff}$ and allowed $\rm \log g$ to vary and a second one in which we fixed $\rm \log g$ and allowed $T_{\rm eff}$ to vary. Both sets of dispersion spectra for six distinct wavelength ranges found to contain features with high degrees of sensitivity changes in $\rm \log g$ and $T_{eff}$ are presented in Figures~\ref{fig:disp_spectra_logg} and \ref{fig:disp_spectra_temp}, respectively.}
Based upon the temperature and surface gravity sensitivities, we selected the following wavelength regions 0.585-0.600~\micron, 0.760-0.840~\micron, 1.155-1.205~\micron, 1.495-1.525~\micron, 2.150-2.27~\micron, and 2.29-2.34~\micron\ over which to perform our spectral fits. However, small adjustments to the wavelength ranges for each region were made for some sources due to residual telluric features or observational defects. For example, the first two wavelength regions were removed from the fits to the coldest stars due to poor SNRs. These regions contain several strong features that are sensitive to surface gravity and temperature such as the sodium doublet at 0.59 \micron, the TiO bands at $\approx 0.7$ \micron, the strong K, Mg, Ti, and Fe lines between 1.16-1.20 \micron, the Mg and Fe lines at 1.50 \micron, the K line at 1.518 \micron, the Si doublet at 2.178 \micron, the Ti and Si lines at 2.18 \micron, the Na doublet at 2.20 \micron, and the CO bandheads longward of 2.29 \micron\ \citep{rayner2009}.

\begin{deluxetable*}{lcccccc}[h!]
\def\arraystretch{0.8}
\tablecaption{Sample Description and Best-Fit Unspotted Parameters.}\label{tab:singlefits}
\tablehead{
\colhead{Star} & \colhead{Region} & \colhead{SpT} & \colhead{$\rm T_{eff}$ (K)} & \colhead{$\rm T_{eff}$ (K)} & \colhead{$\log \rm g$} & \colhead{Rotation Period (d)}}
\startdata
RXJ0445.8+1556 & Taurus & G5 & $5430$ & $5074 \pm 20$ & $3.55 \pm 0.04$ & $1.10$\\
RXJ1508.6-4423 & Lupus & G8 & $5180$ & $5399 \pm 30$ & $4.12 \pm 0.04$ & $0.31$\\
RXJ1526.0-4501 & Lupus & G9 & $5025$ & $5143 \pm 25$ & $4.14 \pm 0.05$ & $2.51$\\
HBC407 & Taurus & K0 & $4870$ & $5091 \pm 30$ & $4.26 \pm 0.05$ & $3.33$\\
PZ99J160843.4-260216 & Upper Scorpius & K0.5 & $4830$ & $4921 \pm 30$ & $3.85 \pm 0.04$ & $2.20$\\
CD-31 12522 & Lupus & K0.5 & $4830$ & $4830 \pm 25$ & $3.82 \pm 0.04$ & $1.99$\\
RXJ1515.8-3331 & Lupus & K0.5 & $4830$ & $4948 \pm 20$ & $4.13 \pm 0.05$ & $2.27$\\
PZ99J160550.5-253313 & Upper Scorpius & K1 & $4790$ & $4926 \pm 15$ & $4.18 \pm 0.03$ & N$\rm A^1$\\
RXJ0457.5+2014 & Taurus & K1 & $4790$ & $4728 \pm 50$ & $3.86 \pm 0.14$ & $1.46$\\
RXJ0438.6+1546 & Taurus & K2 & $4710$ & $4661 \pm 25$ & $3.72 \pm 0.05$ & $3.09$\\
RXJ1608.9-3905 & Lupus & K2 & $4710$ & $4638 \pm 20$ & $3.74 \pm 0.03$ & $2.01$\\
MV Lup & Lupus & K2 & $4710$ & $4760 \pm 30$ & $4.13 \pm 0.05$ & $5.17$\\
RXJ1547.7-4018 & Lupus & K3 & $4540$ & $4748 \pm 50$ & $4.18 \pm 0.15$ & $5.71$\\
RXJ1538.6-3916 & Lupus & K4 & $4375$ & $4579 \pm 30$ & $4.10 \pm 0.14$ & $6.59$\\
MT Lup & Lupus & K5.5 & $4163$ & $4153 \pm 15$ & $3.61 \pm 0.02$ & $4.10$\\
2MASSJ15552621-3338232 & Lupus & K6 & $4115$ & $3971 \pm 10$ & $3.38 \pm 0.02$ & $3.80$\\
RXJ1540.7-3756 & Lupus & K6 & $4115$ & $4052 \pm 10$ & $3.66 \pm 0.02$ & $3.39$\\
RXJ1543.1-3920 & Lupus & K6 & $4115$ & $4083 \pm 15$ & $3.58 \pm 0.02$ & $5.37$\\
MX Lup & Lupus & K6 & $4115$ & $4045 \pm 15$ & $3.44 \pm 0.08$ & $2.28$\\
SO879 & $\sigma$ Ori. & K7 & $4020$ & $3846 \pm 10$ & $3.58 \pm 0.02$ & $9.89$\\
TWA6 & TW Hya & K7 & $4020$ & $3797 \pm 10$ & $3.70 \pm 0.02$ & $0.54$\\
CD -36 7429A & TW Hya & K7 & $4020$ & $3878 \pm 10$ & $3.58 \pm 0.02$ & $5.10$\\
RXJ1607.2-3839 & Lupus & K7.5 & $3960$ & $3815 \pm 10$ & $3.43 \pm 0.02$ & $2.45$\\
MW Lup & Lupus & K7.5 & $3960$ & $3865 \pm 10$ & $4.27 \pm 0.02$ & $3.91$\\
NO Lup & Lupus & K7.5 & $3960$ & $3768 \pm 10$ & $3.70 \pm 0.02$ & $2.81$\\
THA15-43 & Lupus & K7.5 & $3960$ & $3382 \pm 10$ & $4.22 \pm 0.02$ & $3.98$\\
Tyc7760283\_1 (TWA 25) & TW Hya & M0 & $3900$ & $3765 \pm 10$ & $4.05 \pm 0.05$ & $5.08$\\
TWA14 & TW Hya & M0.5 & $3810$ & $3582 \pm 10$ & $4.01 \pm 0.02$ & $0.63$\\
THA15-36A & Lupus & M0.5 & $3810$ & $3674 \pm 10$ & $4.18 \pm 0.02$ & $2.11$\\
RXJ1121.3-3447\_app2 & TW Hya & M1 & $3720$ & $3626 \pm 10$ & $4.10 \pm 0.02$ & $5.45$\\
RXJ1121.3-3447\_app1 & TW Hya & M1 & $3720$ & $3571 \pm 10$ & $3.94 \pm 0.02$ & $5.44$\\
THA15-36B & Lupus & M2 & $3560$ & $3421 \pm 10$ & $4.19 \pm 0.01$ & $2.11$\\
CD -29 8887A & TW Hya & M2 & $3560$ & $3525 \pm 10$ & $4.03 \pm 0.01$ & $4.80$\\
CD -36 7429B & TW Hya & M3 & $3410$ & $3292 \pm 10$ & $3.91 \pm 0.02$ & $5.10$\\
TWA15\_app2 & TW Hya & M3 & $3410$ & $3343 \pm 10$ & $3.85 \pm 0.01$ & $0.65$\\
TWA7 & TW Hya & M3 & $3410$ & $3324 \pm 10$ & $3.85 \pm 0.02$ & $5.04$\\
Sz67 & LupusI & M3 & $3410$ & $3268 \pm 10$ & $3.98 \pm 0.05$ & $0.52$\\
RECX-6 & LupusI & M3 & $3410$ & $3374 \pm 10$ & $3.90 \pm 0.02$ & $1.84$\\
TWA15\_app1 & TW Hya & M3.5 & $3300$ & $3377 \pm 10$ & $3.72 \pm 0.01$ & $0.65$\\
Sz94 & Lupus & M4 & $3190$ & $3270 \pm 10$ & $4.08 \pm 0.02$ & $13.92$\\
SO797 & $\sigma$ Ori. & M4.5 & $3085$ & $3024 \pm 15$ & $3.67 \pm 0.02$ & $13.59$\\
SO641 & $\sigma$ Ori. & M5 & $2980$ & $3000 \pm 20$ & $3.36 \pm 0.04$ & N$\rm A^1$\\
Par\_Lup3\_2 & Lupus & M5 & $2980$ & $3012 \pm 10$ & $3.55 \pm 0.05$ & $8.19$\\
2MASSJ16090850-3903430 & LupusIII & M5 & $2980$ & $3000 \pm 25$ & $3.30 \pm 0.04$ & $0.87$\\
SO925 & $\sigma$ Ori. & M5.5 & $2920$ & $2931 \pm 15$ & $3.26 \pm 0.03$ & N$\rm A^1$\\
SO999 & $\sigma$ Ori. & M5.5 & $2920$ & $2964 \pm 15$ & $3.23 \pm 0.05$ & $13.59$\\
V1191Sco & LupusIII & M5.5 & $2920$ & $2919 \pm 15$ & $2.85 \pm 0.04$ & N$\rm A^1$\\
2MASSJ16091713-3927096 & LupusIII & M5.5 & $2920$ & $2919 \pm 25$ & $3.22 \pm 0.07$ & N$\rm A^1$\\
Sz107 & Lupus & M5.5 & $2920$ & $2961 \pm 15$ & $3.06 \pm 0.02$ & $0.72$\\
Par\_Lup3\_1 & Lupus & M6.5 & $2815$ & $2961 \pm 25$ & $3.00 \pm 0.03$ & N$\rm A^1$\\
LM717 & ChaI & M6.5 & $2815$ & $2678 \pm 20$ & $3.01 \pm 0.03$ & N$\rm A^1$\\
J11195652-7504529 & ChaI & M7 & $2770$ & $2626 \pm 15$ & $4.12 \pm 0.05$ & N$\rm A^1$\\
LM601 & ChaI & M7.5 & $2720$ & $2592 \pm 20$ & $3.82 \pm 0.01$ & N$\rm A^1$\\
CHSM17173 & ChaI & M8 & $2670$ & $2570 \pm 25$ & $3.89 \pm 0.02$ & N$\rm A^1$\\
TWA26 & TW Hya & M9 & $2570$ & $2440 \pm 40$ & $4.09 \pm 0.05$ & $0.42$\\
DENIS1245 & TW Hya & M9.5 & $2520$ & $2402 \pm 40$ & $3.90 \pm 0.10$ & N$\rm A^1$\\
\enddata
\vspace{0.1cm}
\tablenotetext{1}{No TESS or K2 photometry available.}
\end{deluxetable*}

This approach is similar to the ROTFIT code \citep[e.g.,][]{Frasca2017}, where the stellar spectrum is matched to a Doppler shifted, rotationally-broadened template simultaneously fitting for $T_{eff}$, log~g, v~sini, and $v_r$. However, our approach differs in a few key areas: 1) We include wavelength regions across the entire spectral range covered by the X-Shooter spectra from {optical} to NIR $K$-band, 2) we do not subtract the continuum, but rather normalize each spectral window separately, allowing for spectral slopes and shapes of molecular absorption bands to remain unaffected. An advantage of this method is that it is {extinction-insensitive}, as all regions that are fit have been normalized around unity, and they are narrow enough to make differences in the strength of features across wavelength from extinction {unimportant}. 

In all cases, the fits were performed over the temperature range $T_{eff}=2000-7000\ K$ and $log~g=2.5-5.5$ using a Markov-Chain Monte-Carlo (MCMC) algorithm with an \textit{emcee} \citep{Foreman-Mackey2013} sampler under the assumption of flat, uninformative priors. {We used a chi-squared likelihood function defined as}

\begin{equation}
\ln p = -\frac{1}{2} \sum_n \left[ 
\frac{(y_{\rm data} - y_{\rm model})}{\sigma^2} + \ln \left( 2\pi \sigma^2 \right) 
\right]
\end{equation}
\noindent
{where $\sigma$ is the data uncertainty, and $y_{\rm data}$ and $y_{\rm model}$ are the data and model values at each pixel, respectively. The MCMC algorithm was run for 5000 steps with a burn in of 2500 steps for a total of 25 walkers. This was sufficient to ensure convergence across all fits performed, as determined by examining the individual chains. An example of a posterior can be seen in Figure~\ref{fig:posterior}.} {The fits were performed only over the regions shown in Figure~\ref{fig:disp_spectra_logg}, with the log likelihood and reduced chi-squared statistics computed using exclusively the points within these regions.} Besides fitting simultaneously for $T_{eff}$, $\log~g$, and radial velocity ($v_r$), we also fit for rotational broadening ($v\sin{i}$) using the code developed by \citet{Carvalho2023rot}. We do not fit for extinction as these sources were chosen from nearby star forming regions with low extinction \citep{Manara2013, Manara2017, Claes2024}.

\begin{figure*}
\centering
\includegraphics[width=1\linewidth]{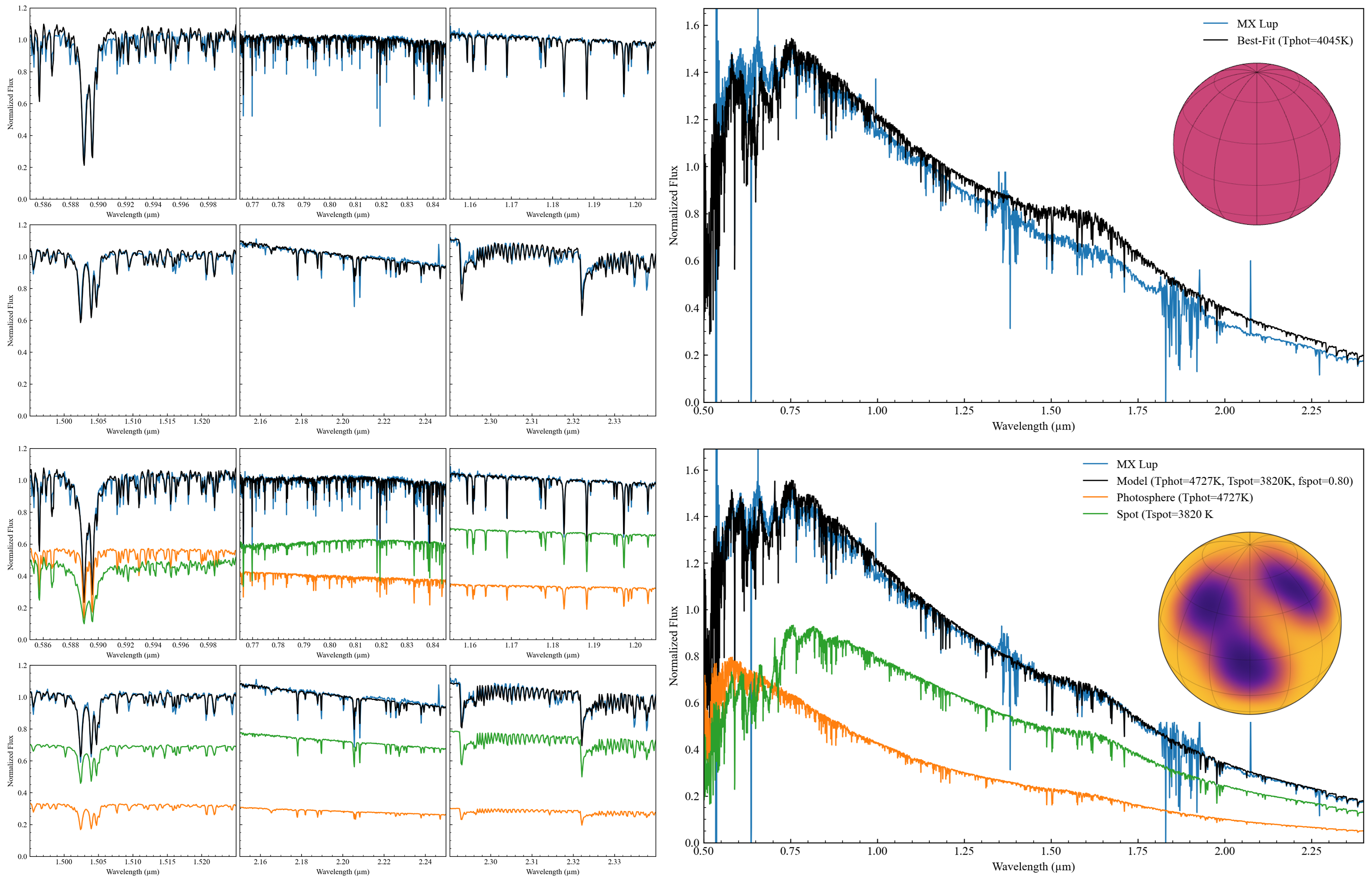}
\caption{\textbf{Comparison between the best-fit unspotted (Top row) and spotted model (Bottom row) for MX~Lup.} In all cases, the stellar spectrum is shown in blue, while the best-fit model is in black. In the spotted models, the starspots are shown in green and the photosphere in orange, where the scale of each component represents their true scaled contribution to the fit. Inset schematic representations of the stellar surface made with \textit{Starry} \citep{Luger2019} are shown for illustrative purposes.}\label{fig:twofits}
\end{figure*}

\begin{figure}
\centering
\includegraphics[width=1\linewidth]{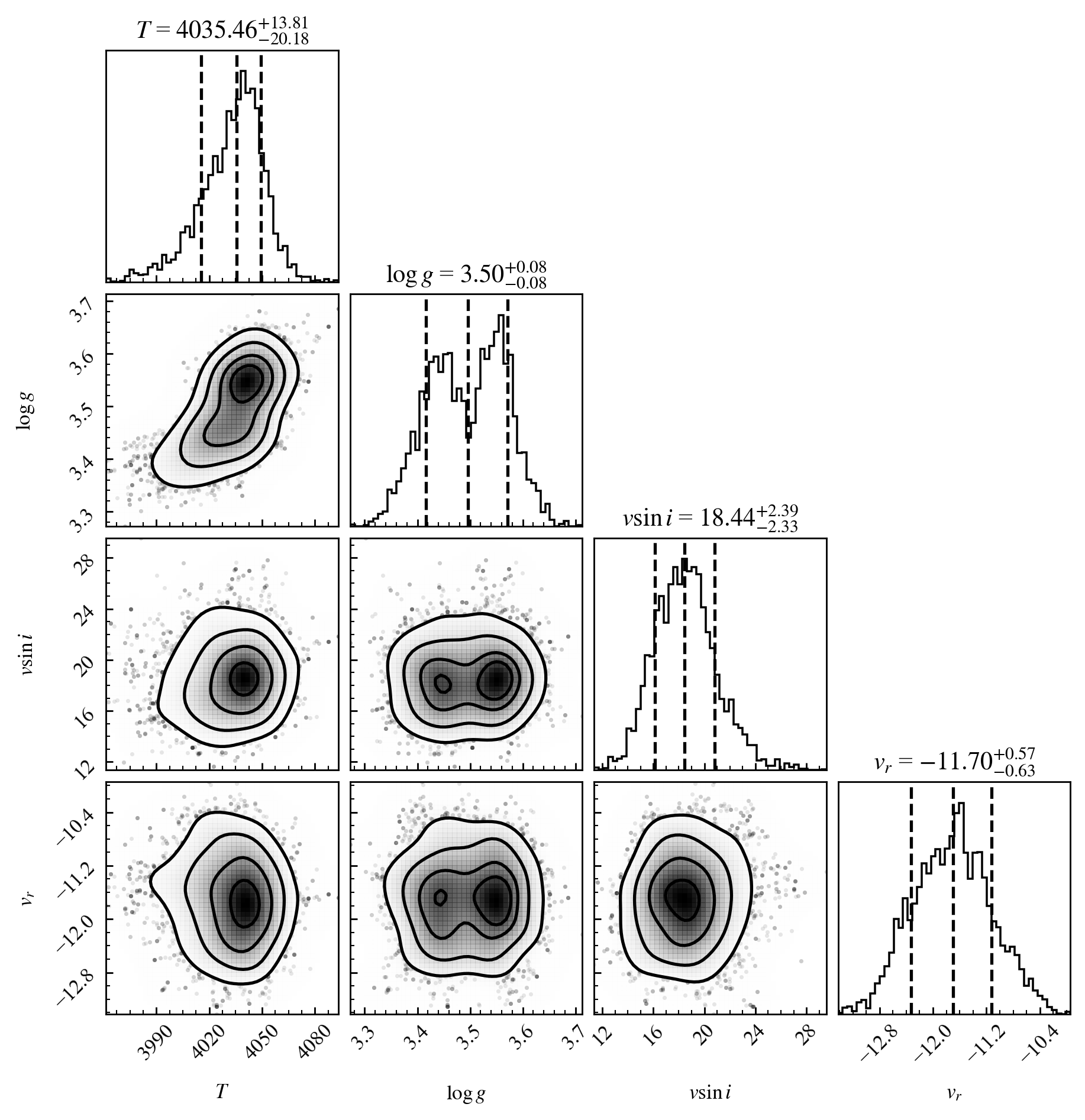}
\caption{\textbf{Posterior for the single-temperature fit in Figure~\ref{fig:twofits}.} Overplotted are the median and one-sigma values.}\label{fig:posterior}
\end{figure}

An example single-temperature fit is shown in Figure~\ref{fig:twofits} (top row) for MX~Lup. By visual inspection, the model does a good job of matching the data, with the strength of the Na doublet at 0.59~\micron\ being matched remarkably well, as are the Mg feature at 1.50~\micron\ and the $K$-band CO bandheads longward of 2.29 \micron. On the other hand, the strength of the TiO bands and the slope are poorly matched in the 0.760-0.840~\micron. Similarly, the strength of the {atomic} Na doublet at 0.82~\micron\ is underestimated. A similar story is seen in the strength of all the strong atomic lines between 1.16-1.20~\micron, the Na doublet at 2.2~\micron, and the Si, Ca, and Fe lines around 2.26~\micron. When we examine the fit across the whole spectrum, we find that the single-temperature fit is unable to reproduce the Spectral Energy Distribution (SED) of the star, with the $H$-band {“hump”}, $K$-band, and everything short of 1~\micron\ being poorly fit.

We used the {single-temperature} fitting routine on all stars present in the \citet{Claes2024} sample, finding very close agreement between our derived temperatures and those quoted therein. Results are shown in column~4 of Table \ref{tab:singlefits}{, and plotted in Figure~\ref{fig:usvthem}.} 

{Note that throughout the manuscript, we use the terms \textit{single-temperature} and \textit{unspotted} interchangeably to describe model fits of stars that lack spots. Whereas \textit{two-temperature} and \textit{spotted} are used, also interchangeably, to refer to model fits of stars that indicate the presence of spots.} 

\subsection{The Issue of Surface Gravity}
Low mass stars contract isothermally during the PMS stage \citep{Hayashi1961} following nearly vertical tracks on HR and Kiel diagrams, allowing for measurements of the stellar luminosities or surface gravities and effective temperatures to yield model-dependent constraints on stellar ages and masses. {Similar to the challenges we have observed for measuring $T_{eff}$ for PMS stars, observational differences in inferred $\rm \log g$ have also been observed.} 

{\citet{Flores2019} placed the star BP~Tau on the Kiel diagram using $\rm \log g$ values and temperatures derived from their fits to high-resolution K-band spectra, as well as values from high-resolution optical spectra found in \citet{Johns-Krull1999}. There is a nearly 0.7 dex difference between the $\rm \log g$ values returned by both methods, as well as a 500~K difference in temperature, resulting in significant age differences regardless of the choice of evolutionary model and isochrones. \citet{Flores2019} attribute this discrepancy to differences in the methods (e.g., optical vs infrared) and details of the evolutionary models. 
In this section, we examine the potential for the enhanced magnetic field strengths in the spots to lead to a greater effective value of $\rm \log g$. While spotted stars certainly have a $\rm \log g$ defined by the gravitational force ($g = GM/R^2$), the spectroscopically-derived $\rm \log g$ may be subject to local pressure variations induced by concentrations of magnetic fields.} 

In the Sun, typical photospheric magnetic fields are on the order of a few tens of Gauss \citep{Schrijver2003}, whereas magnetic field strengths within sunspots reach 3–4~kG, on average, and as high as 7~kG in extreme cases \citep{Siu-Tapia2019}. Zeeman-Doppler Imaging (ZDI) of PMS stars reveal stronger, heterogeneous fields, with kG scale gradients between the photosphere and starspots \citep[e.g.,][]{Finociety2021, Donati2014, Donati2017, Donati2023}. Zeeman-Doppler Imaging (ZDI) of PMS stars reveal highly heterogeneous fields, with kG scale gradients between the photosphere and starspots \citep[e.g.,][]{Donati2014, Donati2017, Donati2023}, hosting strong surface magnetic fields even in unspotted regions of the stellar surface \citep{Finociety2021}. 
Even if PMS stars host strong (~few hundreds of Gauss) surface magnetic fields, the strong 1–3 kG (and possibly much higher) fields measured in starspots suggest that magnetic pressure could easily dominate the thermal gas pressure within the spots. We can estimate the magnetic pressure within a spot as

\begin{equation}
    P_{\rm mag}=\frac{B^2}{8\pi}
\end{equation}
\noindent
where $B$ is the magnetic field strength within the spot. {Following \citet{Calvet1998}, we can obtain an order-of-magnitude estimate of the gas pressure near the depth of continuum formation as}

\begin{equation}
    P_{\rm gas}=\frac{g \tau}{\kappa_{R}}
\end{equation}

\noindent
{where $\tau\approx1$ and $\kappa_{R}$ is the Rosseland mean opacity. This expression provides an analytic approximation of the photospheric pressure and yields values consistent with those tabulated in synthetic atmosphere models for the surface gravities and opacities of typical PMS stars. For this calculation we have used Rosseland mean opacities from \citet{Lederer2009} for solar metallicity.}

If we adopt a representative magnetic field strength of 3~kG measured for sunspots as the strength of the $B$-field in starspots on PMS stars, we find that the gas pressure ($10^4-10^5$~dyne~cm$^{-2}$ for typical opacities and surface gravities of PMS stars) can be overwhelmed by the magnetic pressure of order $4\times10^5$~dyne~cm$^{-2}$. {These pressures are consistent with tabulated values from atmosphere models, which imply equipartition (between magnetic and gas pressure) magnetic field strengths of 1.5-2.5~kG. Observed average magnetic field strengths in PMS stars typically range from 2 to 3~kG \citep{Johns-Krull2007, Shulyak2014}, which is only modestly above equipartition. However, measurements of spot-localized fields reach values of 7-8 kG in some M dwarfs \citep{Shulyak2014}, suggesting that magnetic pressure can significantly exceed thermal gas pressure within the spotted regions. Thus, we theorize that the effect of the magnetic pressure within a spot may force the continuum and spectral features to form deeper in the star and at higher pressure than in the unspotted regions of the photosphere. As a result, the spectral features arising from the spots will yield comparatively higher values $\rm \log g$ and a higher \textit{effective} surface gravity.} 


Another way of describing this would be to note that the opacities in the starspots are lower than in the unspotted photosphere owing to their cooler temperatures \citep{Lederer2009}, and continuum emission will come from deeper in the star where the ambient pressure is higher. Therefore, if we assume that magnetic pressure is negligible within the photosphere, then the emitted spectrum of the starspots will have a surface gravity that can be higher by as much as 1-2~dex compared to the photosphere. 

As was noted by \citet{Safier1999}, pressure equilibrium at the starspot boundary requires a maximum field strength such that equipartition is maintained. However, measurements of field strengths for PMS stars often exceed these limits, as was noted by \citet{Safier1999} and confirmed by magnetic field measurements in the following decades \citep[e.g.,][]{Lopez2019,Flores2021,Flores2022,LopezValdivia2021, Lopez-valdivia2023}. In light of these considerations, it is important to emphasize that surface gravity, as measured by $\rm \log g$ in model fitting is really a statement about the pressure conditions at the depth of spectrum formation, as revealed by pressure broadening of lines and other pressure effects. Therefore, in this case, the measured value of $\rm \log g$ may not be directly related to the mass and radius of the star. At the time, \citet{Safier1999} attributed these inconsistencies to a lack of understanding of the physical processes within stellar atmospheres.

If either of our two theories are correct, that current measurements of magnetic fields on PMS stars are primarily representative of the magnetic fields within spots, and that spectra form at lower temperature and higher pressure within the spot, then we have a likely and reasonable explanation for the discrepant measurements o$\rm \log g$ for PMS stars. 

\begin{deluxetable*}{lcccccccc}
\def\arraystretch{0.8}
\tablecaption{Best-Fit Spotted Parameters.}\label{tab:spotfits}
\tablehead{
\colhead{Star} & \colhead{SpT$^{(1)}$} & \colhead{$\rm T_{eff}^{(1)}$} & \colhead{$\rm f_{ spot}$} & \colhead{$\rm T_{phot}$} & \colhead{$\rm T_{spot}$} & \colhead{$\rm \log g$ (phot)} & \colhead{$\rm \log g$ (spot)} & \colhead{$\rm T_{ eff}$ (spot-corrected)} \\
\colhead{} & \colhead{} & \colhead{(K)} & \colhead{} & \colhead{(K)} & \colhead{(K)} & \colhead{} & \colhead{} & \colhead{(K)}
} 
\startdata
RXJ 0445.8+1556 & G5 & $5430$ & $0.59\pm{0.07}$ & $5966\pm{300}$ & $4168\pm{43}$ & $3.50\pm{0.05}$ & $4.26\pm{0.19}$ & $5139\pm{230}$\\
RXJ 1508.6-4423$^{(2)}$ & G8 & $5180$ & $0.64\pm{0.06}$ & $6936\pm{300}$ & $4242\pm{65}$ & $4.24\pm{0.07}$ & $5.16\pm{0.14}$ & $5674\pm{256}$\\
RXJ 1526.0-4501 & G9 & $5025$ & $0.59\pm{0.03}$ & $5834\pm{85}$ & $3971\pm{70}$ & $4.40\pm{0.05}$ & $5.39\pm{0.09}$ & $4993\pm{81}$\\
HBC 407 & K0 & $4870$ & $0.65\pm{0.05}$ & $5836\pm{250}$ & $3925\pm{50}$ & $4.70\pm{0.08}$ & $5.27\pm{0.15}$ & $3865\pm{182}$\\
PZ99 J160843.4-260216 & K0.5 & $4830$ & $0.63\pm{0.03}$ & $5609\pm{90}$ & $3883\pm{40}$ & $4.29\pm{0.08}$ & $4.84\pm{0.09}$ & $4751\pm{78}$\\
CD-31 12522 & K0.5 & $4830$ & $0.67\pm{0.02}$ & $5488\pm{85}$ & $3717\pm{25}$ & $4.53\pm{0.07}$ & $4.96\pm{0.06}$ & $4546\pm{663}$\\
RXJ 1515.8-3331 & K0.5 & $4830$ & $0.51\pm{0.04}$ & $5278\pm{60}$ & $3822\pm{60}$ & $4.49\pm{0.05}$ & $5.48\pm{0.09}$ & $4702\pm{70}$\\
PZ99 J160550.5-253313 & K1 & $4790$ & $0.53\pm{0.04}$ & $5283\pm{60}$ & $4010\pm{45}$ & $4.37\pm{0.11}$ & $5.21\pm{0.08}$ & $4736\pm{64}$\\
RXJ 0457.5+2014 & K1 & $4790$ & $0.68\pm{0.02}$ & $5113\pm{65}$ & $3426\pm{55}$ & $4.25\pm{0.10}$ & $4.86\pm{0.09}$ & $4205\pm{56}$\\
RXJ 0438.6+1546 & K2 & $4710$ & $0.50\pm{0.03}$ & $4928\pm{30}$ & $3790\pm{30}$ & $3.99\pm{0.05}$ & $5.00\pm{0.10}$ & $4467\pm{39}$\\
RXJ 1608.9-3905 & K2 & $4710$ & $0.49\pm{0.03}$ & $4965\pm{55}$ & $3724\pm{45}$ & $4.14\pm{0.10}$ & $5.06\pm{0.11}$ & $4484\pm{53}$\\
MV Lup & K2 & $4710$ & $0.60\pm{0.03}$ & $5088\pm{60}$ & $3743\pm{35}$ & $4.55\pm{0.07}$ & $5.01\pm{0.05}$ & $4432\pm{56}$\\
RXJ 1547.7-4018 & K3 & $4540$ & $0.54\pm{0.03}$ & $5039\pm{50}$ & $3843\pm{60}$ & $4.50\pm{0.08}$ & $5.38\pm{0.13}$ & $4511\pm{51}$\\
RXJ 1538.6-3916 & K4 & $4375$ & $0.54\pm{0.02}$ & $4848\pm{35}$ & $3776\pm{30}$ & $4.48\pm{0.04}$ & $5.24\pm{0.07}$ & $4368\pm{32}$\\
MT Lup & K5.5 & $4163$ & $0.59\pm{0.01}$ & $4567\pm{30}$ & $3725\pm{25}$ & $4.29\pm{0.09}$ & $5.00\pm{0.04}$ & $4134\pm{22}$\\
2MASS J15552621-3338232 & K6 & $4115$ & $0.74\pm{0.01}$ & $4721\pm{40}$ & $3581\pm{20}$ & $4.68\pm{0.08}$ & $5.00\pm{0.03}$ & $3980\pm{24}$\\
RXJ1540.7-3756 & K6 & $4115$ & $0.72\pm{0.03}$ & $4569\pm{40}$ & $3795\pm{40}$ & $4.32\pm{0.11}$ & $4.72\pm{0.10}$ & $4059\pm{38}$\\
RXJ1543.1-3920 & K6 & $4115$ & $0.69\pm0.02$ & $4634\pm40$ & $3714\pm40$ & $4.41\pm0.09$ & $5.00\pm1.10$ & $4069\pm{34}$\\
MX Lup & K6 & $4115$ & $0.72\pm{0.03}$ & $4661\pm{35}$ & $3730\pm{45}$ & $4.46\pm{0.08}$ & $4.74\pm{0.10}$ & $4059\pm{43}$\\
SO879 & K7 & $4020$ & $0.69\pm{0.05}$ & $4220\pm{40}$ & $3541\pm{30}$ & $3.59\pm{0.10}$ & $4.50\pm{0.06}$ & $3792\pm{44}$\\
TWA6 & K7 & $4020$ & $0.62\pm{0.02}$ & $4117\pm{35}$ & $3293\pm{40}$ & $3.61\pm{0.05}$ & $4.82\pm{0.06}$ & $3673\pm{31}$\\
CD-36 7429A & K7 & $4020$ & $0.75\pm{0.02}$ & $4540\pm{30}$ & $3449\pm{30}$ & $4.36\pm{0.08}$ & $4.88\pm{0.05}$ & $3817\pm{33}$ \\
RXJ1607.2-3839 & K7.5 & $3960$ & $0.79\pm{0.03}$ & $4377\pm{30}$ & $3465\pm{50}$ & $3.73\pm{0.06}$ & $4.65\pm{0.04}$ & $3717\pm{47}$ \\
MWlup & K7.5 & $3960$ & $0.84\pm{0.02}$ & $4569\pm{35}$ & $3669\pm{30}$ & $4.77\pm{0.06}$ & $4.77\pm{0.05}$ & $3860\pm{32}$ \\
NO Lup & K7.5 & $3960$ & $0.84\pm{0.02}$ & $4613\pm{60}$ & $3400\pm{35}$ & $4.62\pm{0.08}$ & $4.64\pm{0.08}$ & $3687\pm{44}$ \\
THA15-43 & K7.5 & $3960$ & $0.87\pm{0.02}$ & $4574\pm{25}$ & $3640\pm{30}$ & $4.71\pm{0.07}$ & $4.74\pm{0.04}$ & $3805\pm{32}$ \\
Tyc7760283\_1$^{(2)}$ (TWA 25) & M0 & $3900$ & $1.00\pm{0.01}$ & $6969\pm{200}$ & $3725\pm{10}$ & $2.51\pm{0.07}$ & $4.36\pm{0.03}$ & $3725\pm{10}$ \\
TWA14 & M0.5 & $3810$ & $0.79\pm{0.01}$ & $4002\pm{20}$ & $3229\pm{20}$ & $3.14\pm{0.06}$ & $4.69\pm{0.04}$ & $3438\pm{17}$ \\
THA15-36A & M0.5 & $3810$ & $0.66\pm{0.02}$ & $3883\pm{25}$ & $3396\pm{10}$ & $3.74\pm{0.04}$ & $4.71\pm{0.04}$ & $3584\pm{16}$ \\
RXJ1121.3-3447\_app2 & M1 & $3720$ & $0.79\pm{0.02}$ & $3954\pm{45}$ & $3370\pm{15}$ & $3.11\pm{0.06}$ & $4.66\pm{0.02}$ & $3518\pm{22}$ \\
RXJ1121.3-3447\_app1 & M1 & $3720$ & $0.68\pm{0.02}$ & $3808\pm{10}$ & $3236\pm{15}$ & $3.00\pm{0.03}$ & $4.74\pm{0.05}$ & $3451\pm{15}$ \\
THA15-36B & M2 & $3560$ & $0.84\pm{0.02}$ & $3850\pm{30}$ & $3101\pm{10}$ & $3.79\pm{0.07}$ & $4.47\pm{0.06}$ & $3259\pm{21}$ \\
CD-29 8887A$^{(2)}$ & M2 & $3560$ & $0.00\pm{0.01}$ & $4743\pm{55}$ & $3389\pm{10}$ & $4.41\pm{0.07}$ & $4.43\pm{0.02}$ & $4743\pm{55}$ \\
CD-36 7429B & M3 & $3410$ & $0.26\pm{0.02}$ & $3366\pm{10}$ & $2969\pm{25}$ & $3.80\pm{0.02}$ & $5.50\pm{0.10}$ & $3276\pm{12}$ \\
TWA15\_app2 & M3 & $3410$ & $0.30\pm{0.01}$ & $3400\pm{100}$ & $2886\pm{10}$ & $3.89\pm{0.02}$ & $5.49\pm{0.01}$ & $3270\pm{79}$ \\
TWA7 & M3 & $3410$ & $0.30\pm{0.02}$ & $3400\pm{10}$ & $2794\pm{35}$ & $3.87\pm{0.02}$ & $5.27\pm{0.10}$ & $3252\pm{15}$ \\
Sz67 & M3 & $3410$ & $0.36\pm{0.03}$ & $3400\pm{10}$ & $2845\pm{35}$ & $3.90\pm{0.04}$ & $5.08\pm{0.12}$ & $3232\pm{19}$ \\
RECX-6 & M3 & $3410$ & $0.19\pm{0.04}$ & $3403\pm{30}$ & $2885\pm{100}$ & $4.00\pm{0.05}$ & $5.21\pm{0.25}$ & $3322\pm{34}$ \\
TWA15\_app1 & M3.5 & $3300$ & $0.60\pm{0.03}$ & $3633\pm{10}$ & $2946\pm{20}$ & $2.97\pm{0.05}$ & $4.97\pm{0.06}$ & $3274\pm{24}$ \\
Sz94 & M4 & $3190$ & $0.32\pm{0.04}$ & $3350\pm{22}$ & $2983\pm{30}$ & $4.00\pm{0.05}$ & $5.42\pm{0.10}$ & $3246\pm{23}$ \\
SO797 & M4.5 & $3085$ & $0.20\pm{0.03}$ & $3049\pm{10}$ & $2718\pm{30}$ & $3.69\pm{0.03}$ & $5.49\pm{0.14}$ & $2991\pm{13}$ \\
SO641 & M5 & $2980$ & $0.28\pm{0.03}$ & $3001\pm{10}$ & $2601\pm{35}$ & $3.59\pm{0.05}$ & $5.33\pm{0.20}$ & $2905\pm{15}$ \\
Par\_Lup3\_2 & M5 & $2980$ & $0.19\pm{0.02}$ & $3020\pm{10}$ & $2727\pm{55}$ & $3.53\pm{0.03}$ & $5.47\pm{0.12}$ & $2971\pm{13}$ \\
2MASSJ16090850-3903430 & M5 & $2980$ & $0.47\pm{0.07}$ & $3034\pm{20}$ & $2756\pm{25}$ & $3.40\pm{0.04}$ & $4.31\pm{0.11}$ & $2913\pm{25}$ \\
SO925 & M5.5 & $2920$ & $0.16\pm{0.03}$ & $2876\pm{15}$ & $2598\pm{50}$ & $3.40\pm{0.03}$ & $5.42\pm{0.08}$ & $2837\pm{16}$ \\
SO999 & M5.5 & $2920$ & $0.18\pm{0.03}$ & $2899\pm{10}$ & $2650\pm{55}$ & $3.39\pm{0.03}$ & $5.47\pm{0.07}$ & $2859\pm{13}$ \\
V1191Sco & M5.5 & $2920$ & $0.53\pm{0.10}$ & $2923\pm{110}$ & $2704\pm{25}$ & $3.59\pm{0.09}$ & $4.00\pm{0.11}$ & $2813\pm{63}$ \\
2MASSJ16091713-3927096 & M5.5 & $2920$ & $0.77\pm{0.01}$ & $3100\pm{10}$ & $2726\pm{10}$ & $3.21\pm{0.05}$ & $4.53\pm{0.04}$ & $2826\pm{9}$ \\
Sz107 & M5.5 & $2920$ & $0.27\pm{0.03}$ & $2908\pm{15}$ & $2526\pm{30}$ & $3.29\pm{0.02}$ & $5.17\pm{0.15}$ & $2820\pm{17}$ \\
Par\_Lup3\_1 & M6.5 & $2815$ & $0.18\pm{0.02}$ & $2621\pm{15}$ & $2516\pm{20}$ & $2.98\pm{0.03}$ & $5.49\pm{0.05}$ & $2603\pm{13}$ \\
LM717 & M6.5 & $2815$ & $0.13\pm{0.03}$ & $2629\pm{15}$ & $2394\pm{40}$ & $3.16\pm{0.03}$ & $5.49\pm{0.08}$ & $2602\pm{15}$ \\
J11195652-7504529 & M7 & $2770$ & $0.41\pm{0.02}$ & $2879\pm{15}$ & $2582\pm{15}$ & $2.50\pm{0.01}$ & $5.49\pm{0.01}$ & $2769\pm{13}$ \\
LM601 & M7.5 & $2720$ & $0.20\pm{0.01}$ & $2580\pm{15}$ & $2564\pm{15}$ & $3.00\pm{0.03}$ & $5.49\pm{0.10}$ & $2577\pm{12}$ \\
CHSM17173 & M8 & $2670$ & $0.37\pm{0.10}$ & $2569\pm{40}$ & $2527\pm{40}$ & $4.12\pm{0.10}$ & $4.14\pm{0.10}$ & $2554\pm{30}$ \\
TWA26 & M9 & $2570$ & $0.20\pm{0.01}$ & $2400\pm{10}$ & $2398\pm{10}$ & $3.26\pm{0.02}$ & $5.43\pm{0.07}$ & $2400\pm{8}$ \\
DENIS1245$^{(2)}$ & M9.5 & $2520$ & $0.00\pm{0.01}$ & $2400\pm{15}$ & $2371\pm{15}$ & $3.43\pm{0.05}$ & $5.49\pm{0.04}$ & $2371\pm{13}$ \\
\enddata 
\vspace{0.1cm}
\tablenotetext{1}{From \citet{Claes2024}.}
\tablenotetext{2}{Unspotted star.}
\end{deluxetable*}

As a means of exploring these scenarios, we updated our spotted star models from \citet{PerezPaolino2024,PérezPaolino2025} to allow for different surface gravities between the photosphere and the spots {and use these models to refit the X-Shooter sample of stars}. We describe this process and results below.  

\subsection{Spotted Models}\label{sec:spotfits}

We follow \citet{PerezPaolino2024} and construct models of spotted stars of the form 

\begin{equation}
\begin{split}
    F_{\lambda,\ \rm model} = F_{\lambda}(T_{\rm phot},\ \log \rm g_{\rm phot})(1-f_{\rm spot}) \\  +\ F_{\lambda} (T_{\rm spot},\ \log \rm g_{\rm spot})f_{\rm spot}
\end{split}
\end{equation}
\noindent
where $F_{\lambda}(T_{\rm phot},\ \log \rm g_{\rm phot})(1-f_{\rm spot})$ represents the contribution from the stellar photosphere, and $F_{\lambda} (T_{\rm spot},\ \log \rm g_{\rm spot})f_{\rm spot}$ represents the contribution from the starspots weighted by the filling factor, $\rm f_{spot}$. We fit models of this form using the same \textit{emcee} MCMC algorithm as in the unspotted case, simultaneously constraining $\rm T_{phot}$, $\rm T_{\rm spot}$, $\rm f_{\rm spot}$, $\rm v\ \rm sini$, $\rm \log g$, and $\rm v_{\rm r}$. {We use the same number of walkers and steps, and check to ensure convergence on each fit as before.}

An example of a spotted fit to the source MX~Lup is shown in the bottom row of Figure~\ref{fig:twofits}. The spotted model (black) does a much better job of matching the molecular and atomic features while capturing the SED of the star than the unspotted fit (top row). Our results suggest that MX~Lup is a heavily spotted ($f_{\rm spot}=0.80$) star with a hot ($T_{\rm phot}=4727\rm\ K$) photosphere and cool ($T_{\rm spot}=3820\ \rm K$) starspots. We present results for fits to all stars in Table \ref{tab:spotfits}.

We correct the effective temperature of a spotted star following \citet{Gully2017} using

\begin{equation}\label{eq:spteff}
    T_{\rm eff}=[T_{\rm phot}^4(1-f_{\rm spot})+T_{\rm spot}^4f_{\rm spot}]^{0.25}. 
\end{equation}
\noindent
For MX~Lup, the spot-corrected effective temperature is {4059~K}, much colder than either the best-fit photospheric temperature of 4727~K or that inferred from its optical spectral type K6~$\approx$ 4115~K \citep{Claes2024}. Their value is close to 4045~K returned by our best-fit single-temperature model found in the previous section. This similarity demonstrates an important point: A spotted star and an unspotted star can have the same or very similar effective temperatures while having wildly discrepant spectra. Both reduced chi-square ($\chi_{\rm red}^2$) and Root-Mean-Square-Error (RMSE) statistics demonstrate that the spotted model with $\chi^2_{red}=0.68$ and RMSE=0.026 is far superior than the unspotted model with $\chi^2_{\rm red}=0.98$ and RMSE=0.032. As can be seen in the spotted model of Figure~\ref{fig:twofits}, the flux from the photosphere dominates the stellar spectrum at wavelengths shorter than $\approx 0.7$\ \micron\, while the spotted regions contribute twice as much flux longward of this wavelength. In total, the starspots contribute 63\% of the total stellar luminosity, while the photosphere contributes only 37\%. 

{We note that the values shown in Figure~\ref{fig:twofits} and in Figure~\ref{fig:posterior} are not identical, as those reported in the text and in the tables are the maximum likelihood values, whereas those in the corner plot are median values. In practice, all of these are within a couple percent of each other.}

In the case of MX~Lup even though unspotted and spot-corrected effective temperatures are nearly identical the spectra are very dissimilar. In general, even the best, most statistically-sound estimates returning singular effective temperatures are prone to systematic errors whereby spectra of the best-fit single temperatures do not reproduce the spectrum of the star at all regions simultaneously. In this regard, the two-temperature, spotted star models provide a significant improvement.

In the case of the MX~Lup fit shown in Figure~\ref{fig:twofits}, we show that the updated fits return a photospheric $\rm \log g$ (phot)~=~4.46 while the starspots return a higher $\rm \log g$ (spot)~=~4.74. Across the entire sample of 56 stars, only three were best fit by unspotted models where the filling factors are either zero or unity. For the spotted stars, all but two were best fit by models in which the spot contributions possessed surface gravities higher than those of the unspotted photospheric components with typical differences on the order of 1~dex. For both stars that did not fit this trend, the bests fit models returned similar values of $\rm \log g$ for the spots and photosphere. In Table~\ref{tab:spotfits} we present the results from the updated spotted star model fits and report $\rm \log g$ values for both the photosphere and the spot.

{We perform a series of tests of the sensitivity of our fitting routine for low SNRs and temperature contrasts between the spots and photosphere, finding that our method is robust down to temperature contrasts of 150~K and SNRs on the order of $\approx$10. Broadly, the detectability of spot signatures depends on four main factors: (1) the temperature contrast between the spot and the unspotted photosphere, (2) the spot filling factor, (3) the spectral wavelength coverage (especially redward of 0.9\micron), and (4) the SNR of the data. Among spectral features, broad molecular absorption bands (e.g., TiO, VO) remain more robust under decreasing SNR than narrow atomic lines, due to their greater equivalent widths and larger breadth across the SED.}

{Three of the stars we analyzed yielded two-temperature fits consistent with an unspotted photosphere, while one additional fit was unsuccessful. RXJ~1508.6$-$4423 returned an unrealistically high photospheric temperature ($T_{\rm phot} = 6936$ K), which we attribute to its high projected rotational velocity ($v\sin i \approx 120$\,km\,s$^{-1}$), consistent with its short rotational period ($P_{\rm rot} = 0.31$ d). DENIS~1245 yielded photospheric and spot temperatures within 29~K of each other. As the coldest star in our sample and the one with the lowest SNR, it is not surprising that it appears to be an outlier. TYC~7760283-1 (TWA~25) also returned an unspotted solution, characterized by an unconstrained photospheric temperature and a spot filling factor near unity. However, Zeeman-Doppler Imaging results from \citet{Nicholson2021} indicate the presence of a large spot on its surface, corroborated by strong rotational modulation in its light curve. The origin of this discrepancy remains unclear. It is possible that our fit was limited by SNR, or that the X-Shooter observation was taken at a point in time when the large singular spot was out of view. We thus consider this result inconclusive. A similar case was observed for CD$-$29~8887A, for which the best-fit model suggests an unspotted photosphere, despite its light curve exhibiting clear rotational modulation \citep{Ricker2015}. However, in this case no ZDI constraints are available for comparison.}

\subsection{Temperature Trends in Starspots}
\begin{figure}
\centering
\includegraphics[width=1\linewidth]{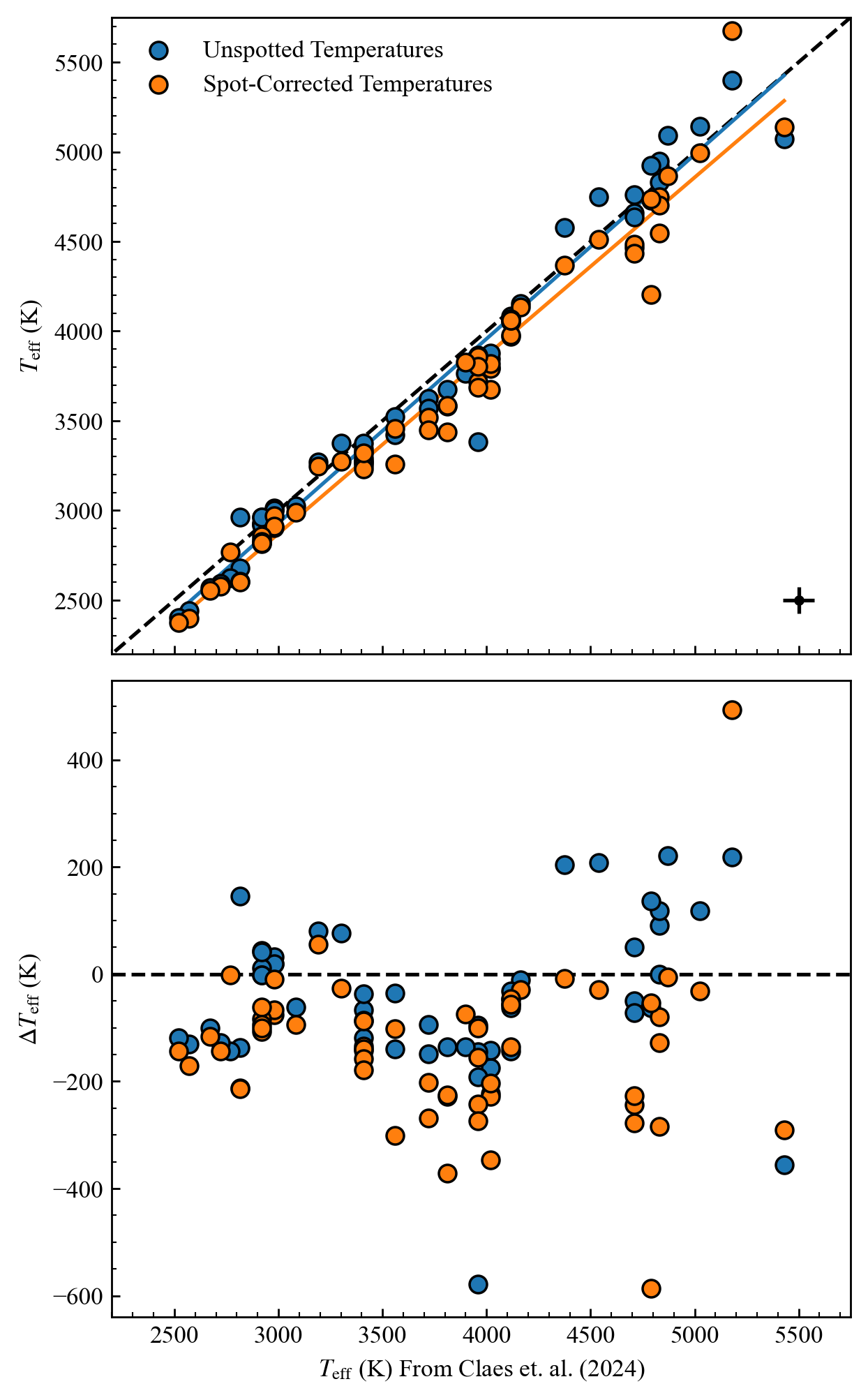}
\caption{\textbf{Comparison between our best-fit spotted (blue) and unspotted (orange) temperatures and those of \citet{Claes2024}.} Overlaid are best-fit lines to indicate trends and systematic differences. The bottom plot shows the difference between our estimates and theirs {($T_{\rm ours}-T_{\rm Claes}$) }with the same color convention.}\label{fig:usvthem}
\end{figure}

In Figure~\ref{fig:usvthem} we show a comparison between the effective temperatures reported in \citet{Claes2024} and both our unspotted (blue dots; Table~\ref{tab:singlefits}, column~5) and spot-corrected (orange dots; Table~\ref{tab:spotfits}, column~9) effective temperatures. Across the entire sample, we find individual temperature differences as large as $\approx500$~K. For temperatures $\rm T_{eff}\geq4500~K$, we see that our fits are $\approx$50~K hotter on average. We attribute these trends to our inclusion of NIR regions in our fitting, whereas \citet{Claes2024} only used optical regions for their analysis. Closer inspection of the residuals (Bottom panel; Figure~\ref{fig:usvthem}) reveals a systematic difference where stars with effective temperatures between 3500-4250~K are colder in the unspotted fits. We attribute this effect to the presence of significant ($\geq 50\%$) starspot coverage in stars of this temperature range, with the impact of starspots decreasing for colder and hotter stars. Spotted fits, on the other hand, are colder on average by $\approx200$~K at all temperatures, with individual differences as large as 600~K. 

\begin{figure}[h]
\centering
\includegraphics[width=1\linewidth]{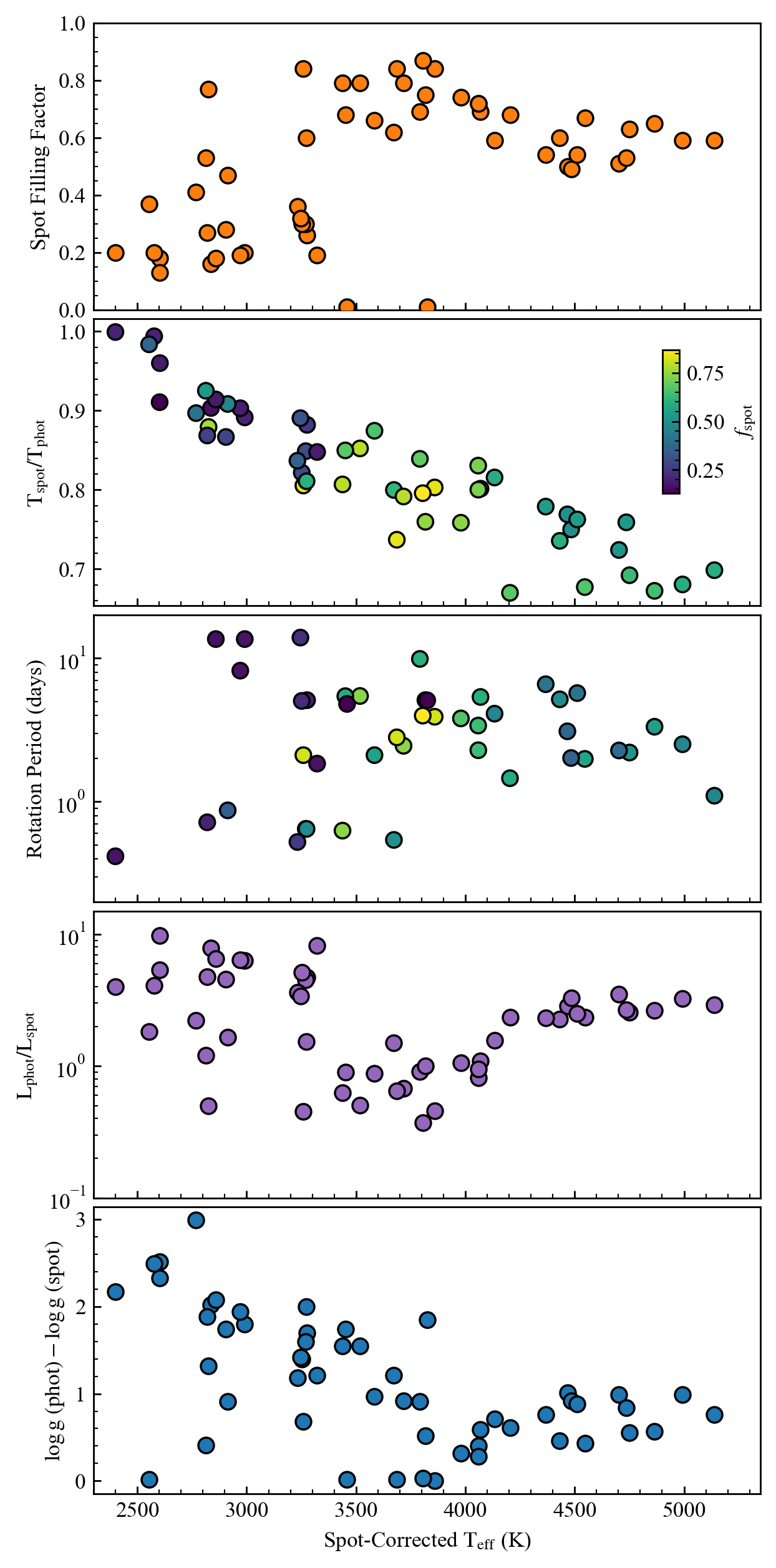}
\caption{\textbf{Starspot Trends Across Temperature.} {Plots of a) starspot filling factors across temperature for stars in our sample, b) ratio of the starspot temperature to the photospheric temperature, colored in by $f_{\rm spot}$. c) rotation periods for stars in the sample colored by spot filling factor, d) ratio of integrated photospheric to starspot luminosity, and d) difference between the starspot surface gravity and the photospheric surface gravity.}} \label{fig:trends}
\end{figure}

In the top panel of Figure~\ref{fig:trends} we plot the spot filling factor returned by our fits across all stars in the sample as a function of spot-corrected effective temperature. A trend becomes evident, whereby stars fall in one of three regimes: 1) Spot coverage for stars with $T_{\rm eff}>4000$~K remains relatively constant with a median value of $0.59\pm0.07$. 2) For stars with $3350\ \rm{K} < T_{\rm eff}<4000$~K the median value of the spot filling factor increases to $0.75\pm0.23$, albeit with a notably larger spread. 3) For stars with $T_{\rm eff} < 3250$~K the coverage decreases as a function of temperature with a median value of $0.29\pm0.22$. Therefore, we separate the behavior of starspots for stars in this sample into three regions based on spot-corrected effective temperature: 1) A saturated region of stars with $T_{\rm eff}>4000$~K, 2) a region of enhanced starspot coverage for stars with $3350\ \rm{K} < T_{\rm eff}<4000$~K, and 3) a region of decreasing starspot coverage correlated with temperature for stars with $T_{\rm eff} < 3250$~K. {However, we caution that the reliability of inferred spot coverage among the coolest stars ($T_{\rm eff}\leq3250$~K) may also be affected by the limited temperature sensitivity of key molecular features at these low temperatures. For example, the TiO bands, which dominate much of the NIR cool star spectrum, exhibit reduced sensitivity to changes in $T_\mathrm{eff}$ below 3200~K \citep{rayner2003}, potentially diminishing the constraining power of our two-temperature fits in this regime. While our fitting windows include a combination of both temperature- and gravity-sensitive lines, as well as more neutral continuum regions, the diagnostic leverage of these features weakens toward the coolest stars. We therefore treat the apparent decline in spot coverage at the lowest temperatures with caution, and consider it alongside the possibility of reduced fitting sensitivity.}

These results are consistent with those of \citet{Gangi}, who found that the offset in effective temperature between those derived using optical and infrared temperatures of young stars is highest for stars with temperatures between 3800--4500~K, with the differences being as large as 500~K. Our result, demonstrating that these stars exhibit a larger degree of starspot activity, supports the conclusion that the discrepancy between optical and near-IR temperatures stems from starspots contaminating the red-optical spectral region of cooler PMS stars. 

Figure~\ref{fig:trends} also shows the ratio of the photospheric temperature to the spot temperature, $T_{\rm spot}/T_{\rm phot}$, as a function of effective temperature. A well-defined linear correlation is found, with the ratio decreasing for higher effective temperatures. Although the trend matches that found by \citet{Herbst2020} and \citet{Berdyugina2005}, those studies included a spot ratio measurement for only one PMS star. The trend found here follows that seen for Main-Sequence dwarfs, but with a much higher temperature difference at the same effective temperature. Our measurements of spot-to-photosphere temperature ratios agree well with those found for smaller samples of stars \citep[e.g.,][]{Gully2017, Gangi, PerezPaolino2024}. On the hotter end, the temperature ratio closely matches that measured for the sun. For late M-dwarfs the temperature difference decreases for colder temperatures, until disappearing fully around 2500~K. This behavior occurs as starspot coverage also decreases for these systems, indicating a general decrease in starspot activity. However, we cannot conclude whether this is representative of a real trend in the behavior of the magnetic topology of these systems or the detectability limits of our models at these cold temperatures. 

Cool M- and L-dwarf atmospheres likely form clouds, which complicates spectroscopic fitting and would render our models uncertain in this regime \citep{allard2014}. However, even the coldest stars in our sample are too warm to form clouds and should experience small discrepancies from cloud formation. {While all but four stellar photospheric temperatures in our sample are above 2600~K, we note that in eight cases the inferred spot temperatures fall below this threshold. These fits may be subject to increased model uncertainty due to the onset of dust formation and limitations in molecular opacities at very low temperatures, and we therefore interpret them with additional caution.}

M-dwarfs with temperatures below $T\approx\ 3350$~K are fully convective, which coincides with where we see a shift in spot filling factors. It is perhaps not surprising that differences in magnetic activity would manifest near this boundary between non-fully convective and fully convective stars. The break we observe may signal a shift in dynamo processes between those operating in PMS stars with an interface between a convective and radiative zone and those which are fully convective \citep{Gregory2012}.
 
In Figure~\ref{fig:trends} we also show rotation periods measured from TESS and K2 photometry of stars in the sample, finding that those with significant starspot coverage ($\geq 30\%$ of the stellar surface) have rotation periods between 0.2 and 10 days and temperatures over 3250~K. However, we see no clear correlation between rotation period, temperature, and starspot coverage. All of the stars in our sample are in the saturated regime of activity relations given their short periods. Therefore, we do not expect their starspot characteristics to be modulated by rotation, and attribute the scatter in this plot to sample selection biases and random chance. 

The luminosity of a spotted star depends on the temperature of the starspots and photosphere, as well as the starspot filling factor. Therefore, we computed the ratio of the luminosity of the starspots to the photospheric luminosity for stars in our sample assuming simple blackbody curves and show the results in Figure~\ref{fig:trends}. For stars in the high-temperature, constant starspot regime and in cold, fully convective stars, the luminosity of the photosphere is greater than that of the spots. However, for stars in the region of enhanced starspot activity the luminosity of the photosphere can be as little as 30\% of the starspot luminosity, indicating that studies estimating bolometric corrections from single-temperature analysis of stars in this range should cautiously interpret such estimates. 

Surface gravity differences between the photosphere and starspots are also shown in Figure~\ref{fig:trends}. For stars with $T_{\rm eff} > 4000$~K, the $\log g$ difference remains roughly constant, with a median value of $0.71 \pm 0.22$~dex. Below 4000~K, however, the difference increases with decreasing temperature, reaching values as large as 2~dex in the coldest stars. This trend suggests a stronger pressure differential between spot and photosphere in cooler stars, potentially indicating that the spectra of the spots form in the deeper, denser magnetically confined regions of the spots.

{We note that part of the elevated $\log g$ values inferred for the spotted components may partially reflect limitations in the synthetic spectra, particularly in reproducing molecular features such as TiO in cool, low-gravity atmospheres. In this regime, higher surface gravity values may act as compensatory parameters to improve spectral fits, especially in the coolest stars where the spot temperature approaches that of late M dwarfs. Regardless, we tested the influence of $\log g$ on the temperatures derived by the fitting routine, finding differences under 100~K.}

In light of these surface gravity trends, a consistent picture emerges for spotted PMS stars. The spectral contributions from the cool spots are significant, particularly in the infrared, affecting the integrated fluxes, the global effective temperatures, and the values of $\rm \log\ g$. Commonly used single-temperature models systematically misrepresent the fundamental parameters of spotted PMS: 1) effective temperatures are consistently overestimated from optical spectra, 2) SEDs, particularly in the NIR, are poorly fit, and 3) surface gravity values also depend on the spectral region used for the analysis. The surface gravity values recovered from the flux of the spotted regions is often 1-2 dex larger than the values derived from the unspotted photospheric component. Ignoring the contributions of the spots to the overall surface gravity on spotted PMS stars will lead to values of $\rm \log\ g$ that are not physically representative of the stellar masses and radii. These effects are strongest in the 3300–3700~K range, where spot flux dominates the red-optical and NIR continuum, but persist across the full temperature span of our sample. Importantly, the success of two-temperature fits suggests that no single $\rm T_{eff}$ or $\rm \log\ g$ meaningfully characterizes a spotted PMS star.

\section{The Magnetic Evolution of Pre-Main Sequence Stars}
{Since less massive, and therefore colder, M-dwarfs display stronger magnetic fields \citep{Shulyak2014}, one would expect the starspot filling factor to increase for lower temperatures. However, this is the opposite of what we find. This seemingly counterintuitive result can be interpreted as a consequence of the differing magnetic topologies of M dwarfs across the mass spectrum. If predominantly non-axisymmetric magnetic configurations are to blame for the filling factors we measure being on average lower for colder stars, it would also explain the large range of filling factors for fully convective stars. \citet{Gregory2012} use ZDI magnetic topology maps for a sample of PMS stars to study the effects of radiative core development on magnetic topology across the PMS. Their results suggest that the large-scale field topology of PMS stars is strongly dependent on the internal stellar structure, with the development of a radiative core. The development of a tachocline is suggested to coincide with a period of magnetic reconfiguration from primarily axisymmetric dipolar fields with magnetic energy concentrated on large scale structures to highly complex, mostly non-axisymmetric fields with magnetic energy concentrated on smaller scales. Even though the specifics of the physical mechanism responsible for this evolution are unknown, there is strong evidence that the loss of a primarily dipolar field coincides with the development of a radiative core. While this result is inferred from ZDI maps, \citet{Saunders2009} using I-band photometry of stars in the $\approx$~13 Myr \textit{h Per} found that the fraction of periodically variable TTSs decreases for stars on the radiative core side of the convective boundary. The authors attribute this to a change in the spot configurations and propose that the spots of low-mass, fully convective stars are dominated by larger polar spots, as opposed to smaller scale spots on stars with radiative cores. 
}

{ZDI studies have found a simultaneous trend in magnetic topology and field strength on both sides of the fully convective boundary, with lower mass M-dwarfs having stronger, preferentially axisymmetric poloidal fields, and higher mass M-dwarfs having weaker, more non-axisymmetric fields in line with theoretical predictions \citep{Morin2010Late, Morin2008Mid, Donati2008Early, See2015}. Given the structural similarities between M-dwarfs and PMS stars, it is reasonable to propose a similar behavior in PMS stars.
}

\begin{figure*}[htb]
\centering
\includegraphics[width=0.8\linewidth]{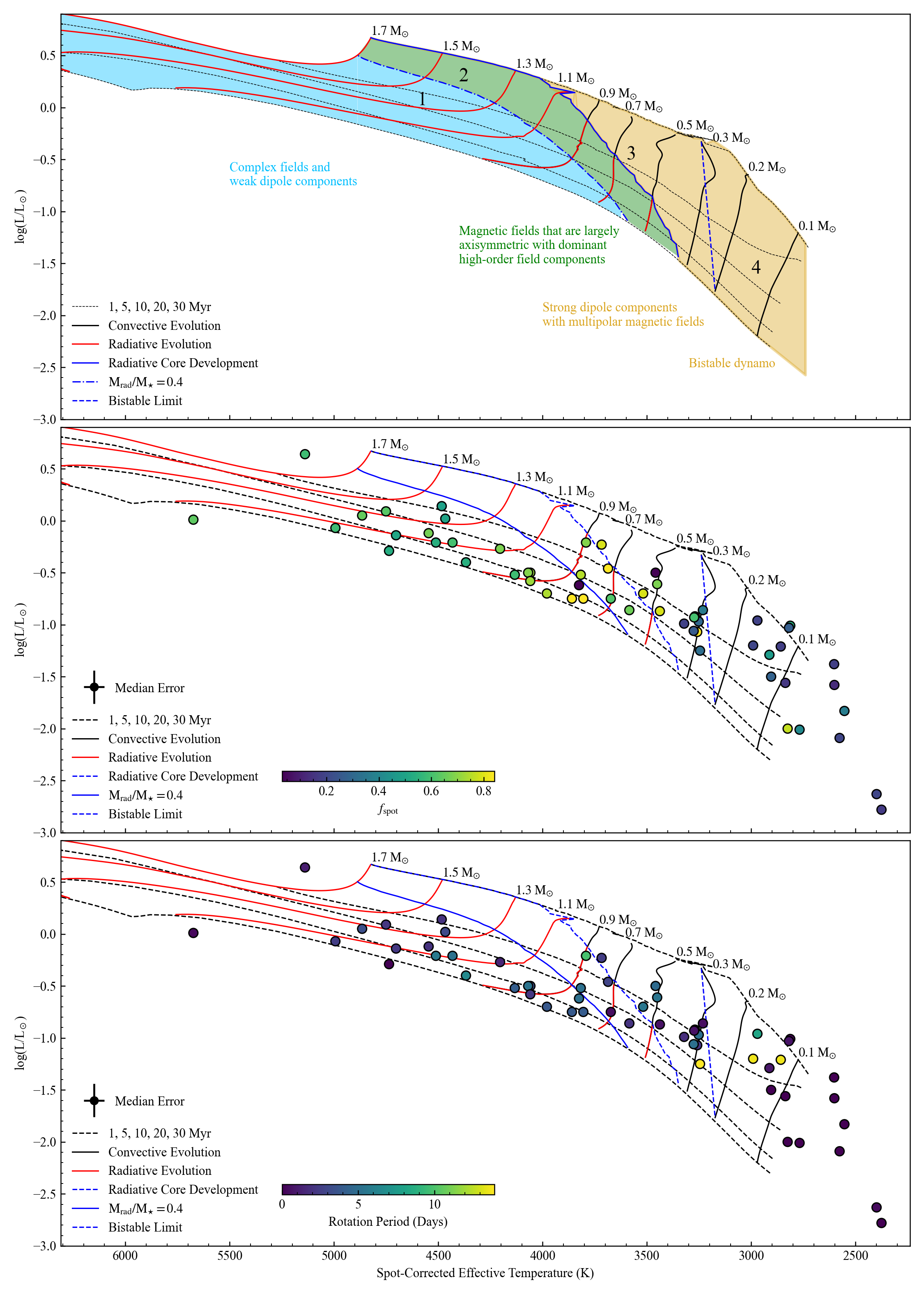}
\caption{\textbf{Magnetic Topology Across the HR Diagram.} The top panel shows theoretical regimes of magnetic topology from \citet{Gregory2012} overlaid on evolutionary tracks (solid {black and red} lines) and isochrones (dashed {black}lines) from the \citet{Feiden2016} magnetic models. Colored background regions indicate distinct magnetic configurations: complex weak dipoles (blue), axisymmetric fields with high-order components (green), strong dipole-dominated multipolar fields (orange), and the bistable dynamo regime (tan). {The Solid blue and brown line indicate the onset of radiative core development and constant $\rm M_{\rm rad}/M_\star = 0.4$, respectively, while the dashed blue line indicates the start of the bistable regime.} The middle panel plots our Class III PMS sample, color-coded by spot coverage ($f_{\rm spot}$). Spot coverage peaks near the radiative transition zone and declines toward both higher- and lower-mass regimes, consistent with distinct magnetic behavior across structural regimes. The bottom panel shows the same stellar sample color-coded by rotation period. The lack of a strong correlation between rotation and location on the diagram reflects that all stars lie within the saturated regime of magnetic activity, where rotation no longer governs spot emergence.}
\label{fig:MagneticHR}\end{figure*}

{Based on this observations, \citet{Gregory2012} propose an HR-Diagram-based classification scheme for the magnetic behavior of stars, which we show in Figure~\ref{fig:MagneticHR}. Stars with substantial radiative cores ($\rm M_{rad}/M_{\star}\geq0.4$) host complex fields with weak dipole components. {Stars with small radiative cores ($\rm 0\leq M_{rad}/M_{\star}\leq0.4$) have large axisymmetric fields with magnetic energy concentrated in small regions.} Fully convective stars that are close to the fully convective boundary host strong dipole components in their multi-polar fields, with those below some previously never observed boundary hosting stars with simple axisymmetric fields with strong dipole components as well stars with complex non-axisymmetric fields with weak dipole components. {Within this bistable regime are stars that host a variety of field topologies, some with dipolar axisymmetric field components and others with non-axisymmetric complex fields hosring weak dipole components}. These results were expanded by \citet{See2015} with a factor of five larger sample of stars across a broader age range, largely confirming the conclusions of \citet{Gregory2012} while adding a further layer of complexity. Their results suggest that the latitude of starspot formation is higher for faster rotators. With access to ZDI maps from different observing seasons, \citet{See2015} finds that stars in the bistable and weak radiative regimes can display large degrees of magnetic topology variability on timescales of months and years. Across their sample, the authors find that stars less massive than $\rm M_{\star}\leq0.5\ M_{\odot}$ tend to have small toroidal energy fractions compared to stars with $\rm M_{\star}\geq0.5\ M_{\odot}$. To first order, these results appear to be insensitive to the strength of differential rotation in low-mass stars.
}

{We schematically showcase these regions in the top panel of Figure~\ref{fig:MagneticHR} using the \citet{Feiden2016} magnetic evolutionary models (Hereon referred to as F16) evolutionary models, where a dashed blue line marks the point of radiative core development, and solid blue lines mark the points where a star reaches $\rm M_{rad}/M_\star=0.4$. In the middle panel we plot all of the stars in our sample colored by $\rm f_{spot}$ alongside evolutionary tracks and isochrones from F16. The reasons for this choice of model will be discussed in the following section. Inspection of the middle panel reveals that the region of enhanced starspot activity (maximum starspot coverage) overlaps with the region of the HR Diagram contained between radiative core development and the formation of a moderately strong ($M_{\rm rad}/M_\star=0.4$) radiative core. Stars to the left of this boundary (more massive and hotter at a given age) host lower starspot filling factors with small star-to-star variations. Stars less massive than the line of radiative core formation are fully convective, and tend to have smaller starspot filling factors. However, a region spanning roughly $\rm 0.2\ M_{\odot}$ to the right of the fully convective boundary seem to include both moderately and weakly spotted stars. These results within uncertainties seem to follow the predictions of \citet{Gregory2012}, a result that is unsurprising when we consider that a measurement of the surface starspot coverage from spectroscopy is a form of measurement of the magnetic topology of the star. 
}
{In the bottom panel of Figure~\ref{fig:MagneticHR} we plot the same comparison but now with stars colored according to their rotation period. We find no evidence for a correlation or trend across the sample, a result consistent with our assertion that these stars are all in the saturated regime of stellar activity trends. These results suggest that, for fully saturated stars, rotation plays no discernible role in determining their surface magnetic topology at these young ages and low masses. 
}

{Our results complement the X-ray study of \citet{Argiroffi2016}, who analyzed magnetic activity regimes in the $\approx$13 Myr-old h Persei cluster. Their identification of saturated and supersaturated X-ray activity, alongside emerging activity–rotation correlations, marks the onset of main-sequence-like magnetic behavior, which they attribute to the transition from turbulent to $\alpha\Omega$ dynamo operation. In our sample of non-accreting T Tauri stars spanning a broader age range, we observe a steep change in starspot morphology near the fully convective boundary, suggesting that this structural transition, when stars develop radiative cores, imprints strongly on surface magnetic topologies. While they study the coronal (X-ray) manifestation of magnetic evolution, our data provide the photospheric counterpart, revealing a strong mass dependence in spot morphology and coverage that tracks stellar interior evolution. Together, these results support a unified picture where the emergence of radiative cores catalyze a fundamental shift in both the generation and surface expression of magnetic fields during the PMS.}
 
{The emergence of radiative cores may very well mark a pivotal shift in magnetic behavior, from large-scale, axisymmetric fields to increasingly complex and variable topologies - a shift mirrored in the distribution of surface starspots across the HR diagram. Our findings support this picture as the starspot coverage peaks near the onset of radiative core development and declines on either side, consistent with expectations from both dynamo simulations and ZDI observations. Yet the interpretation of these trends, and their implications for stellar masses and ages, depends critically on the choice of evolutionary models. In particular, models that account for the structural and thermal effects of magnetic fields are required to correctly place these stars on theoretical tracks. In the following section, we compare several models that incorporate starspots and magnetic fields to clearly illustrate this point.
}

\section{Inferring Fundamental Stellar Parameters of Spotted PMS Stars}
Deriving accurate masses and ages for pre-main-sequence stars is a central goal in stellar astrophysics, yet one complicated by the presence of strong magnetic fields and widespread surface inhomogeneities such as starspots. These effects distort the fundamental relation between luminosity, effective temperature, and stellar structure, shifting stars on the HR diagram away from the positions predicted by standard, non-magnetic models \citep[e.g.,][]{Gully2017,myself,PerezPaolino2024}. To address this, recent generations of evolutionary models have incorporated magnetic inhibition of convection and spot-induced flux suppression, enabling {a more theoretically robust} treatment of magnetically active young stars. In this section, we assess how these different physical treatments affect the inferred stellar parameters of our sample.

\subsection{Placing Spotted Stars on the HR Diagram}
We place each object in our sample on a HR Diagram using both non spot-corrected and spot-corrected effective temperatures. For this, we have used stellar luminosities from \citet{Claes2024} who used wider, low resolution slits to obtain accurate flux calibration followed by narrower, higher resolution slits to obtain the spectra analyzed here. This method provides well-calibrated spectroscopy, with median uncertainties in {$\rm log\ {(L/L_{\odot})}$} on the order of 0.16 dex. Given that the luminosities were computed from the broad-coverage X-Shooter spectra with extrapolations performed only in the UV and IR spectral ranges, we believe that these luminosities include the contributions of spots to the luminosity of the star.

Both the F16 and SPOTS models introduce the effects of magnetism into their calculations of stellar structure, albeit through different mechanisms. The F16 models are a modified version of the magnetic Dartmouth stellar evolution models \citep{Feiden2012}, modeling the interaction of magnetic fields with convection and modifying the plasma equation of state by coupling Maxwells equations to the equations of stellar structure. These models assume equipartition magnetic field strengths at the stellar surface, with values in line with observations of 1-3 kG. {The radial profile of the average magnetic field strengths scale following a gaussian profile, peaking at interior magnetic fields of 50 kG for models with surface $\langle Bf \rangle\approx2.5$ kG}. Compared to non-magnetic evolutionary models \citep[e.g.,][]{Baraffe2015} the presence of strong magnetic fields blocks convection, resulting in tracks and isochrones that are shifted to cooler temperatures and higher luminosities, giving stars larger radii and luminosity at a given age. 

\begin{figure*}[htb]
\centering
\includegraphics[width=1\linewidth]{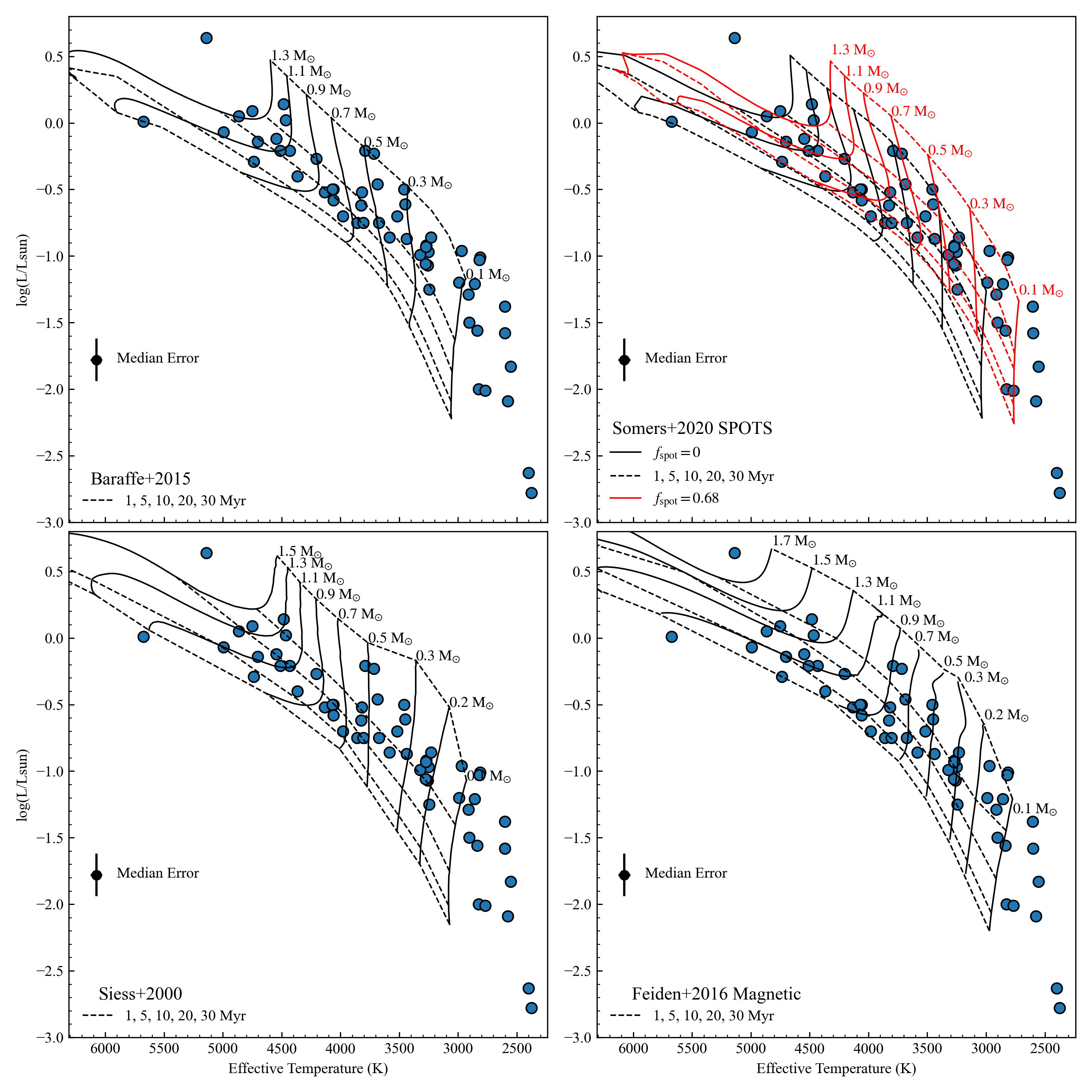}
\caption{\textbf{Comparison of the positions of spotted Class III pre-main-sequence stars on the Hertzsprung–Russell diagram against several evolutionary model grids.} All points represent spot-corrected effective temperatures and luminosities derived from our spectral modeling. Models shown include \citet[][top left]{Baraffe2015}, \citet[][SPOTS models, top right]{Somers2015}, \citet[][bottom left]{Siess2000}, and \citet[][F16 magnetic models, lower right]{Feiden2016}. Solid black lines denote mass tracks and dashed black lines show isochrones at 1, 5, 10, 20, and 30 Myr. In the Somers+2020 panel, red lines indicate evolutionary tracks for a spot filling factor of $f_{\rm spot}=0.54$, highlighting the displacement of tracks relative to the unspotted ($f_{\rm spot}=0$) case in black.} \label{fig:HR_comp}
\end{figure*}

The SPOTS models also introduce the effects of magnetism but through the incorporation of the flux blocking effect of starspots on the stellar surface and their effect on the surface boundary conditions. These have the effect of altering the pressure and temperature at the stellar photosphere, and to suppress convective energy transport throughout the stellar interior. Like F16, the SPOTS models assume pressure equilibrium (i.e., equipartition) between the starspots and the stellar photosphere. The net effect of starspots on the evolutionary tracks of stars is to shift stars at a given mass and age toward lower temperatures and higher luminosities, slowing their contraction along the Hayashi tracks, with the magnitude of this shift being proportional to the starspot filling factor. Testing demonstrates that the F16 models are closest to the SPOTS models when $\rm f_{spot}\approx0.54$, suggesting that the treatment of magnetic fields, whether by direct incorporation as is the case of F16 or through implementing the convection-blocking effect through starspots in the SPOTS models, provides similar results. The effects of magnetic/starspot effects on stellar evolution are shown in Figure~\ref{fig:HR_comp}, where we have placed our sample of PMS stars alongside the traditional evolutionary models of \citet[][Top left]{Baraffe2015} and \citet[][Lower left]{Siess2000} as well as the magnetic models of \citet[][Lower right]{Feiden2016} and the spotted models of \citet[][Top right]{Somers2020} for two starspot filling factors of $\rm f_{spot}=0$ (black lines) and $\rm f_{spot} = 0.54$ (red lines). 

In order to calculate masses and ages for stars in our sample, we interpolate the evolutionary tracks and isochrones across temperature and luminosity and use a MCMC algorithm to simultaneously determine masses and ages, as well as uncertainties. For the SPOTS models, the grids are interpolated in $\rm f_{spot}$ as well as mass and age, since every star has a different starspot filling factor that needs to be accounted for in the models, and by default these are computed only for $\rm f_{spot}=0,\ 0.17,\ 0.31,\ 0.51,\ 0.68,\ and\ 0.85$. {We have computed unspotted masses using the best-fit single-temperatures and the $f_{\rm spot}=0$ SPOTS models as well as the non-magnetic F16 models, seeing very little difference in the masses inferred between the two.} We then computed ages and masses with the F16 magnetic models and the SPOTS models at the corresponding starspot filling factor for each star. These values are reported in Table \ref{tab:masses}, where the upper and lower uncertainties correspond to $\rm 1\sigma$ deviations in the posteriors.

\begin{deluxetable*}{ccccccccc}[htb]
\tablecaption{Ages and Masses of PMS Stars.}\label{tab:masses}
\def\arraystretch{0.85}
\tablehead{
\colhead{} & \multicolumn{4}{c}{Unspotted $\rm T_{eff}$} & \multicolumn{4}{c}{Spot-Corrected $\rm T_{eff}$} \\
Model & \multicolumn{2}{c}{SPOTS ($\rm f_{spot} =0$)} & \multicolumn{2}{c}{F16 \textbf{(Non-Magnetic)}}  & \multicolumn{2}{c}{SPOTS ($\rm f_{spot} = \rm var$)} & \multicolumn{2}{c}{F16 (Magnetic)} \\
\colhead{Star} & \colhead{$\rm M/M_{\odot}$} & \colhead{Age (Myr)} & \colhead{$\rm M/M_{\odot}$} & \colhead{Age (Myr)} & \colhead{$\rm M/M_{\odot}$} & \colhead{Age (Myr)} & \colhead{$\rm M/M_{\odot}$} & \colhead{Age (Myr)}}
\startdata
RXJ0445.8+1556 & \nodata & \nodata & \nodata & \nodata & \nodata & \nodata & $1.25\pm0.13$ & $0.76^{+0.50}_{-0.30}$ \\
RXJ1508.6-4423 & $1.09\pm{0.06}$ & $21.42^{+5.73}_{-4.09}$ & $1.16\pm{0.07}$ & $21.17^{+16.28}_{-5.71}$ & \nodata & \nodata & $1.07\pm0.06$ & $35.15^{+7.59}_{-7.63}$ \\
RXJ1526.0-4501 & $1.11\pm{0.06}$ & $19.14^{+7.57}_{-3.84}$ & $1.14\pm{0.08}$ & $28.63^{+13.63}_{-9.27}$ & $1.09\pm0.05$ & $27.86^{+12.96}_{-7.32}$ & $1.11\pm0.08$ & $23.64^{+7.20}_{-5.91}$ \\  
HBC407 & $1.25\pm{0.07}$ & $11.21^{+7.94}_{-3.67}$ & $1.30\pm{0.12}$ & $14.66^{+16.88}_{-6.12}$ & $1.20\pm0.08$ & $15.97^{+9.44}_{-6.29}$ & $1.19\pm0.09$ & $15.25^{+9.07}_{-6.02}$ \\  
PZ99J160843.4-260216 & $1.3\pm{0.03}$ & $7.32^{+1.8}_{-1.49}$ & $1.28\pm{0.06}$ & $9.03^{+16.07}_{-3.00}$ & $1.27\pm0.05$ & $11.41^{+3.00}_{-2.73}$ & $1.24\pm0.07$ & $9.04^{+3.15}_{-2.31}$ \\
CD-31 12522 & $1.15\pm{0.05}$ & $13.06^{+6.14}_{-2.92}$ & $1.11\pm{0.06}$ & $15.22^{+18.87}_{-5.03}$ & $1.14\pm0.06$ & $14.64^{+7.71}_{-3.49}$ & $1.16\pm0.09$ & $12.78^{+4.33}_{-3.04}$ \\
RXJ1515.8-3331 & $1.10\pm{0.05}$ & $16.50^{+4.24}_{-3.05}$ & $1.10\pm{0.06}$ & $17.73^{+16.73}_{-4.58}$ & $1.11\pm0.05$ & $18.84^{+12.80}_{-5.07}$ & $1.11\pm0.08$ & $17.95^{+6.00}_{-4.38}$ \\
PZ99J160550.5-253313 & $0.97\pm{0.05}$ & $26.70^{+12.39}_{-5.18}$ & $0.98\pm{0.05}$ & $24.91^{+8.66}_{-5.12}$ & $1.01\pm0.05$ & $31.58^{+11.67}_{-7.99}$ & $0.98\pm0.06$ & $31.79^{+8.57}_{-7.10}$ \\
RXJ0457.5+2014 & $1.03\pm{0.05}$ & $17.42^{+4.72}_{-3.53}$ & $1.00\pm{0.06}$ & $15.85^{+6.34}_{-5.13}$ & $1.06\pm0.06$ & $12.66^{+6.37}_{-3.41}$ & $1.08\pm0.06$ & $11.45^{+3.74}_{-2.63}$ \\
RXJ0438.6+1546 & $1.19\pm{0.08}$ & $6.12^{+5.77}_{-1.47}$ & $1.11\pm{0.07}$ & $4.51^{+12.73}_{-2.11}$ & $1.18\pm0.06$ & $6.11^{+2.08}_{-1.31}$ & $1.28\pm0.11$ & $6.87^{+2.09}_{-1.69}$ \\
RXJ1608.9-3905 & $1.20\pm{0.06}$ & $4.02^{+3.23}_{-1.15}$ & $1.13\pm{0.07}$ & $2.81^{+5.52}_{-0.88}$ & $1.22\pm0.04$ & $4.09^{+1.17}_{-0.89}$ & $1.22\pm0.13$ & $4.19^{+1.56}_{-1.18}$ \\
MV Lup & $1.08\pm{0.14}$ & $14.42^{+12.63}_{-7.05}$ & $1.08\pm{0.11}$ & $16.71^{+17.99}_{-9.42}$ & $1.10\pm0.10$ & $13.18^{+15.07}_{-6.81}$ & $1.09\pm0.12$ & $15.02^{+13.58}_{-7.08}$ \\
RXJ1547.7-4018 & $1.07\pm{0.05}$ & $14.64^{+4.17}_{-3.02}$ & $1.04\pm{0.06}$ & $12.32^{+11.41}_{-5.66}$ & $1.06\pm0.06$ & $17.20^{+13.06}_{-4.91}$ & $1.10\pm0.08$ & $16.89^{+5.88}_{-4.03}$ \\
RXJ1538.6-3916 & $0.94\pm{0.04}$ & $21.86^{+13.50}_{-5.48}$ & $0.90\pm{0.06}$ & $19.25^{+10.02}_{-6.08}$ & $0.97\pm0.06$ & $25.68^{+11.77}_{-6.87}$ & $0.99\pm0.06$ & $25.82^{+8.66}_{-5.75}$ \\
MT Lup & $0.82\pm{0.05}$ & $11.60^{+16.06}_{-6.55}$ & $0.78\pm{0.05}$ & $10.77^{+11.34}_{-6.26}$ & $0.93\pm0.09$ & $20.45^{+15.75}_{-10.72}$ & $0.96\pm0.10$ & $20.28^{+14.69}_{-9.65}$ \\
2MASSJ15552621-3338232 & $0.70\pm{0.04}$ & $16.43^{+12.26}_{-4.03}$ & $0.72\pm{0.04}$ & $13.12^{+5.45}_{-3.11}$ & $0.83\pm0.04$ & $37.83^{+7.30}_{-7.01}$ & $0.82\pm0.03$ & $29.92^{+8.04}_{-5.60}$ \\
RXJ1540.7-3756 & $0.76\pm{0.04}$ & $8.25^{+1.92}_{-1.55}$ & $0.75\pm{0.04}$ & $8.05^{+3.88}_{-1.74}$ & $0.94\pm0.04$ & $21.02^{+5.48}_{-4.10}$ & $0.94\pm0.04$ & $18.01^{+4.36}_{-3.27}$ \\
RXJ1543.1-3920 & $0.78\pm{0.05}$ & $9.13^{+2.58}_{-2.04}$ & $0.76\pm{0.05}$ & $8.44^{+9.43}_{-2.82}$ & $0.94\pm0.05$ & $21.08^{+6.59}_{-4.71}$ & $0.94\pm0.04$ & $18.43^{+4.95}_{-3.72}$ \\
MX Lup & $0.75\pm{0.04}$ & $12.04^{+2.74}_{-2.30}$ & $0.76\pm{0.04}$ & $10.69^{+9.69}_{-3.09}$ & $0.90\pm0.04$ & $28.01^{+6.71}_{-5.20}$ & $0.89\pm0.03$ & $23.93^{+5.91}_{-4.52}$ \\
SO879 & $0.56\pm{0.03}$ & $1.94^{+1.70}_{-0.69}$ & $0.52\pm{0.04}$ & $1.41^{+1.48}_{-0.67}$ & $0.82\pm0.06$ & $3.26^{+4.32}_{-1.61}$ & $0.88\pm0.09$ & $3.08^{+3.12}_{-1.65}$ \\
TWA6 & $0.58\pm{0.02}$ & $9.57^{+11.77}_{-5.14}$ & $0.59\pm{0.03}$ & $9.18^{+12.22}_{-5.13}$ & $0.72\pm0.05$ & $16.02^{+16.89}_{-9.01}$ & $0.69\pm0.04$ & $12.66^{+11.60}_{-6.17}$ \\
CD-36 7429A & $0.62\pm{0.03}$ & $5.32^{+6.56}_{-2.79}$ & $0.62\pm{0.04}$ & $4.76^{+6.83}_{-2.59}$ & $0.84\pm0.06$ & $12.26^{+13.40}_{-6.81}$ & $0.84\pm0.06$ & $9.68^{+10.18}_{-4.84}$ \\
RXJ1607.2-3839 & $0.53\pm{0.01}$ & $1.44^{+0.28}_{-0.23}$ & $0.50\pm{0.01}$ & $1.39^{+0.28}_{-0.27}$ & $0.80\pm0.05$ & $3.04^{+0.95}_{-0.69}$ & $0.81\pm0.06$ & $2.63^{+0.67}_{-0.55}$ \\
MW lup & $0.63\pm{0.03}$ & $12.72^{+3.06}_{-2.47}$ & $0.65\pm{0.03}$ & $11.86^{+8.91}_{-2.98}$ & $0.77\pm0.04$ & $34.17^{+7.49}_{-6.25}$ & $0.78\pm0.03$ & $25.05^{+6.25}_{-4.71}$ \\
NO Lup & $0.53\pm{0.03}$ & $3.23^{+3.34}_{-1.48}$ & $0.52\pm{0.04}$ & $2.78^{+3.60}_{-1.44}$ & $0.81\pm0.06$ & $7.50^{+9.84}_{-4.17}$ & $0.74\pm0.06$ & $5.13^{+5.10}_{-2.53}$ \\
THA15-43 & $0.29\pm{0.01}$ & $2.34^{+0.58}_{-0.51}$ & $0.29\pm{0.03}$ & $2.93^{+2.78}_{-0.49}$ & $0.78\pm0.04$ & $31.08^{+7.02}_{-5.77}$ & $0.76\pm0.03$ & $21.71^{+7.30}_{-4.48}$ \\
Tyc7760283 1 (TWA 25) & $0.56\pm{0.03}$ & $5.54^{+6.23}_{-2.83}$ & $0.55\pm{0.04}$ & $5.17^{+7.24}_{-2.82}$ & $0.60\pm0.07$ & $6.92^{+10.04}_{-3.87}$ & $0.81\pm0.09$ & $13.06^{+13.56}_{-6.81}$ \\
TWA14 & $0.43\pm{0.01}$ & $7.76^{+9.26}_{-4.00}$ & $0.44\pm{0.04}$ & $6.89^{+7.64}_{-3.45}$ & $0.65\pm0.04$ & $16.14^{+16.70}_{-9.08}$ & $0.46\pm0.03$ & $7.62^{+6.45}_{-3.50}$ \\
THA15-36A & $0.50\pm{0.02}$ & $9.63^{+10.41}_{-4.94}$ & $0.51\pm{0.03}$ & $9.05^{+10.51}_{-4.76}$ & $0.69\pm0.05$ & $18.39^{+16.40}_{-9.86}$ & $0.60\pm0.04$ & $13.27^{+12.17}_{-6.41}$ \\
RXJ1121.3-3447 app2 & $0.45\pm{0.02}$ & $4.59^{+5.06}_{-2.35}$ & $0.45\pm{0.03}$ & $4.46^{+4.66}_{-2.2}$ & $0.71\pm0.05$ & $11.40^{+13.80}_{-6.42}$ & $0.55\pm0.05$ & $6.29^{+6.03}_{-3.07}$ \\
RXJ1121.3-3447 app1 & $0.41\pm{0.02}$ & $3.26^{+3.01}_{-1.41}$ & $0.40\pm{0.02}$ & $2.70^{+2.63}_{-1.31}$ & $0.60\pm0.04$ & $5.34^{+7.14}_{-2.82}$ & $0.50\pm0.05$ & $3.66^{+3.16}_{-1.75}$ \\
THA15-36B & $0.33\pm{0.01}$ & $8.59^{+12.92}_{-4.88}$ & $0.31\pm{0.03}$ & $7.41^{+11.20}_{-5.14}$ & $0.55\pm0.03$ & $19.30^{+16.90}_{-10.66}$ & $0.30\pm0.02$ & $6.46^{+4.74}_{-2.61}$ \\
CD-29 8887A & $0.38\pm{0.01}$ & $2.02^{+1.71}_{-0.70}$ & $0.36\pm{0.02}$ & $1.67^{+1.50}_{-0.97}$ & $0.35\pm0.02$ & $2.03^{+1.52}_{-0.73}$ & $0.52\pm0.06$ & $2.68^{+2.37}_{-1.35}$ \\
CD-36 7429B & $0.24\pm{0.01}$ & $4.47^{+3.47}_{-2.03}$ & $0.24\pm{0.03}$ & $7.52^{+7.86}_{-3.98}$ & $0.32\pm0.02$ & $6.22^{+6.27}_{-3.17}$ & $0.32\pm0.02$ & $2.68^{+2.37}_{-1.35}$ \\
TWA15 app2 & $0.27\pm{0.01}$ & $3.03^{+2.79}_{-1.29}$ & $0.26\pm{0.02}$ & $3.89^{+3.06}_{-1.54}$ & $0.33\pm0.05$ & $4.17^{+5.36}_{-2.26}$ & $0.32\pm0.07$ & $4.55^{+3.92}_{-2.02}$ \\
TWA7 & $0.26\pm{0.01}$ & $3.08^{+2.98}_{-1.40}$ & $0.25\pm{0.02}$ & $3.38^{+3.32}_{-1.75}$ & $0.31\pm0.02$ & $4.07^{+4.47}_{-2.06}$ & $0.30\pm0.02$ & $4.82^{+3.28}_{-1.90}$ \\
Sz67 & $0.23\pm{0.02}$ & $1.73^{+0.38}_{-0.41}$ & $0.24\pm{0.02}$ & $2.52^{+3.62}_{-1.8}$ & $0.32\pm0.02$ & $2.89^{+0.68}_{-0.54}$ & $0.30\pm0.02$ & $3.39^{+0.45}_{-0.30}$ \\
RECX-6 & $0.30\pm{0.01}$ & $4.97^{+3.14}_{-0.86}$ & $0.28\pm{0.03}$ & $5.45^{+1.69}_{-0.85}$ & $0.32\pm0.03$ & $5.31^{+1.66}_{-1.20}$ & $0.36\pm0.03$ & $6.75^{+1.40}_{-1.16}$ \\
TWA15 app1 & $0.30\pm{0.01}$ & $4.47^{+6.70}_{-2.12}$ & $0.28\pm{0.03}$ & $4.81^{+6.04}_{-2.14}$ & $0.45\pm0.03$ & $8.04^{+10.06}_{-4.29}$ & $0.32\pm0.03$ & $4.74^{+3.38}_{-1.93}$ \\
Sz94 & $0.22\pm{0.01}$ & $6.86^{+5.62}_{-3.32}$ & $0.21\pm{0.02}$ & $6.78^{+5.08}_{-2.31}$ & $0.33\pm0.02$ & $11.41^{+12.48}_{-5.98}$ & $0.28\pm0.02$ & $9.92^{+7.28}_{-4.13}$ \\
SO797 & $0.10\pm{0.01}$ & $1.49^{+0.89}_{-0.36}$ & $0.11\pm{0.01}$ & $1.52^{+1.49}_{-0.89}$ & $0.15\pm0.02$ & $1.85^{+1.62}_{-0.74}$ & $0.14\pm0.02$ & $2.75^{+2.72}_{-1.89}$ \\
SO641 & $0.10\pm{0.01}$ & $2.75^{+2.44}_{-1.16}$ & $0.09\pm{0.01}$ & $2.11^{+3.69}_{-2.13}$ & $0.12\pm0.01$ & $3.17^{+3.61}_{-1.60}$ & $0.12\pm0.01$ & $5.58^{+2.31}_{-2.07}$ \\
Par Lup3 2 & \nodata & \nodata & $0.14\pm{0.02}$ & $0.66^{+0.98}_{-0.40}$ & $0.11\pm{0.01}$ & $0.47^{+0.76}_{-0.27}$ & $0.15\pm0.02$ & $1.03^{+2.14}_{-0.77}$ \\
2MASSJ16090850-3903430 & $0.10\pm{0.01}$ & $1.75^{+1.19}_{-0.55}$ & $0.09\pm{0.01}$ & $2.61^{+2.24}_{-1.74}$ & $0.18\pm0.03$ & $3.07^{+3.53}_{-1.57}$ & $0.12\pm0.01$ & $3.84^{+1.98}_{-2.38}$ \\
SO925 & \nodata & \nodata & $0.12\pm{0.01}$ & $6.85^{+3.45}_{-2.40}$ & $\cdots$ & $\cdots$ & $0.11\pm0.01$ & $4.06^{+1.37}_{-1.62}$ \\
SO999 & \nodata & \nodata & $0.14\pm{0.02}$ & $3.09^{+2.23}_{-2.04}$ & $\cdots$ & $\cdots$ & $0.11\pm0.01$ & $2.46^{+2.04}_{-1.62}$ \\
V1191Sco &  & \nodata & $0.13\pm{0.01}$ & $1.11^{+0.97}_{-0.48}$ & $\cdots$ & $\cdots$ & $0.12\pm0.02$ & $0.78^{+0.61}_{-0.30}$ \\
2MASSJ16091713-3927096 &  &  & $0.11\pm{0.01}$ & $19.09^{+3.59}_{-2.70}$ & $0.23\pm0.01$ & $46.75^{+2.85}_{-4.51}$ & $0.09\pm0.01$ & $18.53^{+2.77}_{-2.89}$ \\
Sz107 & \nodata & \nodata & $0.14\pm{0.02}$ & $1.42^{+2.02}_{-0.96}$ & $\cdots$ & $\cdots$ & $0.11\pm0.01$ & $0.76^{+1.48}_{-0.54}$ \\
Par Lup3 1 & \nodata & \nodata & $0.13\pm{0.01}$ & $4.83^{+2.89}_{-2.67}$ & $\cdots$ & $\cdots$ & $0.18\pm0.04$ & $0.55^{+1.52}_{-0.40}$ \\
LM717 & \nodata & \nodata & $0.12\pm{0.01}$ & $2.44^{+0.52}_{-0.54}$ & $\cdots$ & $\cdots$ & $0.18\pm0.04$ & $1.37^{+3.52}_{-0.92}$ \\
J11195652-7504529 & \nodata & \nodata & $0.12\pm{0.03}$ & $4.55^{+5.32}_{-2.37}$ & $0.10\pm0.00$ & $13.36^{+3.23}_{-2.35}$ & $0.09\pm0.01$ & $19.59^{+2.10}_{-5.99}$ \\
LM601 & \nodata & \nodata & $0.18\pm{0.03}$ & $19.35^{+10.54}_{-11.57}$ & $\cdots$ & $\cdots$ & $0.20\pm0.01$ & $29.18^{+3.96}_{-6.37}$ \\
CHSM17173 & \nodata & \nodata & $0.22\pm{0.03}$ & $14.21^{+2.44}_{-8.95}$ & $\cdots$ & $\cdots$ & $\cdots$ & $\cdots$ \\
TWA26 & \nodata & \nodata & \nodata & \nodata & $\cdots$ & $\cdots$ & $\cdots$ & $\cdots$ \\
DENIS1245 & \nodata & \nodata & \nodata & \nodata & $\cdots$ & $\cdots$ & $\cdots$ & $\cdots$ \\
\enddata
\end{deluxetable*}

\subsection{Spot-Corrected Masses and Ages} \label{sec:hr_comp}
\begin{figure*}[htb]
\centering
\includegraphics[width=1\linewidth]{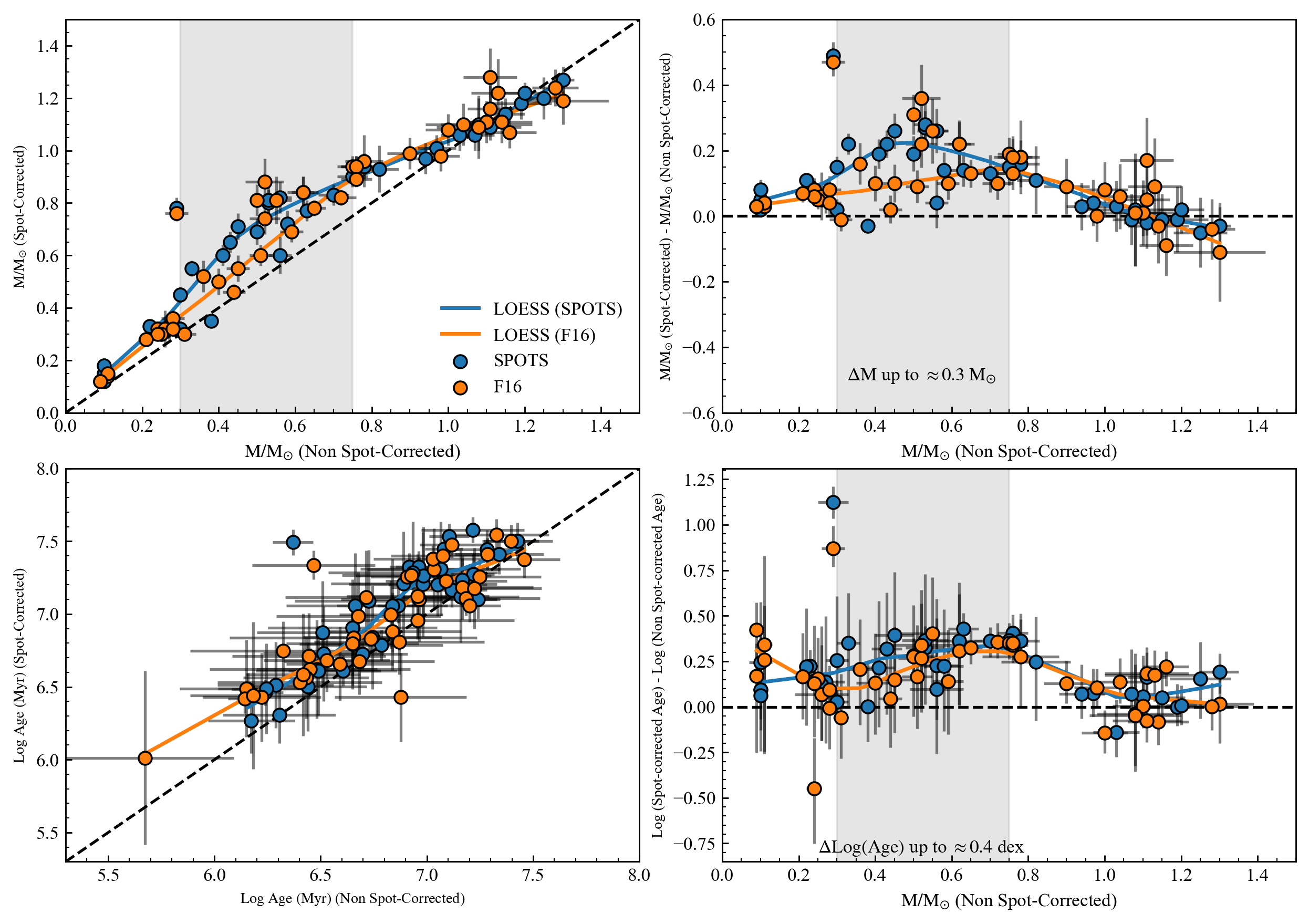}
\caption{\textbf{Masses and Ages Derived from Non Spot-Corrected and Spot-Corrected Analyses.} The top row shows that both the SPOTS and F16 models yield higher masses when spot-corrected $\rm T_{eff}$ are used, with differences reaching up to 0.3 $M_{\odot}$ at $M \leq 0.8 M_{\odot}$. At fixed mass, spot-corrected temperatures also lead to older age estimates (bottom row), partially flattening the mass–age trend commonly observed in young populations. While spot corrections do not remove the age spread entirely, they reduce both the slope and inter-model disagreement, indicating that cool spots are a major driver of apparent youth in low-mass stars. The shaded region highlights the parameter space most affected by surface inhomogeneities. LOESS curves are overplotted in blue and red.} \label{fig:mass_age}
\end{figure*}

Having derived spot-corrected masses and ages, we now assess how these inferences differ depending on the models used. The left column of Figure~\ref{fig:mass_age} shows a comparison between non spot-corrected masses and ages computed using the F16 (orange points) and the SPOTS models (blue points), while the right column shows residuals between the two as a function of non spot-corrected mass. We chose to plot these as a function of the non spot-corrected stellar masses to make it easier to infer the scale of the systematic biases that arise from starspots for stars of a known non spot-corrected, traditionally determined mass.

The main striking trend from Figure~\ref{fig:mass_age} is that biases incurred from starspots are smallest for stars with $\rm M\leq 0.3\ M_{\odot}$ and for stars with $\rm M\geq 0.8\ M_{\odot}$, with masses in the range $\rm 0.3\ M_{\odot} \leq M\leq 0.8\ M_{\odot}$ showing the largest differences across models when applying starspot corrections. We have overplotted LOcally Estimated Scatterplot Smoothing (LOESS) curves to guide the eye in these comparisons \citep{Cleveland1979}.

Maximal differences between non spot-corrected and spot-corrected masses for stars in the $\rm 0.3\ M_{\odot} \leq M\leq 0.8\ M_{\odot}$ are as high as $\Delta \rm M \approx \rm 0.3\ M_{\odot}$, which represents a 80\% difference for a $\rm 0.4\ M_{\odot}$ star. Mass shifts of this magnitude, if uncorrected, could bias the shape of the IMF inferred from HR diagram modeling, especially at the low-mass end. Age differences can be as large as 0.5 dex for $\rm 0.4\ M_{\odot}$ stars. The $\rm 0.3\ M_{\odot} \leq M\leq 0.8\ M_{\odot}$ range coincides with the mass interval over which young stars develop radiative cores at the ages considered here. This is the region where we find spot coverage to be maximal as stars become partially radiative, and their structural response to surface boundary conditions becomes more pronounced. Accordingly, spot corrections produce the largest shifts in derived parameters precisely where the underlying stellar structure is most susceptible to surface inhomogeneities.

Model-dependent differences are apparent across all mass and age ranges, with the discrepancy between spot-corrected and non-spot-corrected values being significantly smaller when using the magnetic F16 models. In fact, masses and ages inferred from non spot-corrected temperatures using F16 are often very close to those derived with spot-corrected values. Does this imply that the F16 models eliminate the need for explicit spot corrections? We argue otherwise. The F16 models incorporate the effects of magnetic fields by introducing strong, equipartition-strength fields in the stellar interior during the 0.1--0.5 Myr period of stellar evolution. {These magnetic fields are prescribed based on stellar mass and age, and are turned on in the models between $\sim$0.1-0.5 Myr, treating magnetism as a permanent perturbation throughout the rest of the evolution of the star.} In contrast, the SPOTS models account for surface magnetism through starspot coverage, which modifies the boundary conditions and suppresses convection throughout the entire PMS evolution. Because the starspot filling factor can be tuned, this effectively becomes a phenomenological proxy for interior magnetic suppression (i.e.,~$\rm \langle Bf \rangle \propto f_{\rm spot}$). Thus, the F16 models can be viewed as implicitly encoding a fixed spot correction via an interior magnetic field. The result is a natural reduction in the offset between spot-corrected and non-spot-corrected parameters, since the magnetic shift in the F16 HR tracks is comparable in magnitude to the temperature shift from spot corrections. This visual agreement between the F16 spot-corrected and non spot-corrected masses and ages is thus somewhat coincidental and not evidence that spot effects are captured in the F16 physics explicitly, but rather that their magnetic suppression roughly mimics the cooling effect of moderate spot coverage. Critically, this match breaks down outside the limited age/mass regime where F16 magnetic fields are imposed, and does not account for the heterogeneous spot properties seen across young stars. In the future, models of stellar evolution should be developed to account for internal magnetism in the equations of stellar structure while simultaneously introducing the convection blocking effect of starspots and their corresponding modified boundary conditions at the stellar surface. 

We note that spot-corrected masses and ages reduce the mass dependent scatter in the sample by about 0.3 dex. However, the sample of stars is constructed from many star forming regions with different ages, so no further inferences should be made on this point from this dateset. Starspots pose a possible correction, reducing but not entirely eliminating, the mass dependent age trends seen in star forming regions. This 0.3 dex reduction is comparable to the full width of the age spreads typically inferred in individual SFRs, suggesting that a significant portion of this spread may be attributable to spot-induced $\rm T_{eff}$ biases rather than intrinsic age dispersion.

\section{Discussion}\label{sec:discussion}
\subsection{On the Feasibility of Such Large Starspot Filling Factors}
Estimating the surface coverage of starspots on pre-main sequence (PMS) stars remains observationally challenging, with different methods producing seemingly contradictory results. Three primary diagnostics are used: photometric variability modeling, spectroscopic line-depth or continuum fitting, and magnetic field measurements either via Zeeman Doppler Imaging (ZDI) or Zeeman broadening of magnetically sensitive lines.

Each technique probes different aspects of the stellar surface. ZDI is sensitive primarily to large-scale magnetic structures, capturing the geometry and polarity of dominant magnetic regions but missing small-scale fields due to different polarities cancelling \citep{Kochukhov2020}. Zeeman broadening, by contrast, reflects the total unsigned surface-averaged magnetic field strength and is insensitive to magnetic polarity. Consequently, ZDI typically returns field strengths of a few hundred gauss and inferred spot coverage fractions below 30\% \citep[e.g.,][]{Williamo2022}, while Zeeman broadening studies often infer kilogauss-level fields and much higher magnetic filling factors \citep[e.g.,][]{Johns-Krull2007}.

This discrepancy has led to two interpretations: either the stellar surface is globally magnetized with minimal structure, or starspots and magnetic regions are confined to a fraction of the surface, as in the solar case. Given that ZDI only detects the largest coherent magnetic structures, it is best interpreted as a lower bound on spot coverage. Zeeman broadening and spectroscopic modeling, in contrast, offer upper bounds, integrating over both large and small-scale features. We argue that the true starspot filling factor across the entire stellar surface likely lies between these extremes—typically in the 30–60\% range for the stars analyzed here and consistent with results from our two-temperature fits.

Inclination effects further complicate the interpretation. Because the visibility of starspots depends on the stellar orientation, a nearly pole-on star with a large polar spot may appear to have a higher filling factor than it actually does. ZDI has confirmed the presence of polar and near-polar starspots in PMS stars \citep[e.g.,][]{Donati2017, Donati2019}, and MHD simulations show that spot advection toward the poles is a natural outcome of large-scale magnetic dynamo processes \citep{Yadav2015}, as was mentioned above from MHD models. As shown by \citet{Jackson2013}, stars with high spot coverage but many small, evenly distributed spots can exhibit minimal photometric variability, making photometry a poor tracer of total f$_{\rm spot}$. These effects could plausibly account for a few extreme cases of high spot coverage. However, the tight trends in filling factors with temperature observed across our statistically significant sample (Figure~\ref{fig:trends}) cannot be explained by inclination alone. The systematic nature of these trends, especially as a function of temperature or mass, support the conclusion that high spot filling factors often covering the majority of the visible hemisphere are a genuine physical property of young, magnetically active stars.

\subsection{Reconsidering Effective Temperatures and Spectral Types in the Presence of Starspots and Magnetic Fields}
{If the photospheric emission from a PMS is not produced by one mostly isothermal atmosphere, but instead is a synthesis of at least two different temperature components, how might we characterize such a star's effective temperature and/or spectral type? We have observed that spectral types derived from optical indices tend to match the temperature of the unspotted photosphere in the two-temperature models, while NIR indices tend to track the temperature of the cooler component. This has led to the same star being classified as, for example, late K-type from optical lines and mid M-type from infrared spectral indices \citep[e.g.,][]{Gully2017}. In a narrow sense, both classifications are “correct," but neither reflects the full radiative output of the star. Furthermore, this determination of spectral type may be variable in time as the spot configuration on the stellar surface evolves to cover more or less of the star at a given time. Given that optical vs. infrared spectral type mismatches and color index discrepancies for T~Tauri stars have been known issue for decades, it comes as no surprise that a unique spectral type and effective temperature may not exist of such stars as they do for main sequence stars. }


{The spectra of heavily spotted stars in this sample contain molecular and atomic absorption features with strengths corresponding to atmospheres several hundred Kelvin cooler than would be based on their optical spectra, and even infrared spectral types. In some cases, the continuum flux from the spotted photospheric regions contributes more to the IR spectra and significantly alters the shape of the SED. Therefore, line ratios, molecular band depths, or SEDs alone are insufficient to determine accurate effective temperatures or spectral types. Even when used collectively to determine a multi-component fit over both the optical and IR regions, one must still naively assume that the resultant averaged temperature may not be representative of the spectral emission at all wavelengths. Thus, the standard relationship between effective temperature, luminosity, and stellar radius, $\rm T_{eff}=(L_\star/4 \pi R_\star^2 \sigma)^{1/4}$, does not apply. Additionally, the presence of spots may also impact the traditional conversions between spectral type and surface temperature such as those of \citet{Pecaut2013}, which were calibrated on stars assumed to be unspotted or only mildly spotted. In light of the two-temperature models and the difficulty defining a representative effective temperatures, we find that applying the usual one-to-one mapping of SpT–$\rm T_{eff}$ lacks meaning for heavily spotted stars. While we have argued that spot-corrected effective temperatures (Equation~5) provide a more physically meaningful description of the bolometric emission from the multi-component stellar photospheres, even these temperatures do not likely capture the shape of the spectrum outside of the modeled regions. 
}

The sensitivity of commonly used temperature indices to spot coverage implies that the assignment of a single-temperature-based spectral type to a spotted PMS star is not merely uncertain, but perhaps even fundamentally ambiguous. While optical and infrared classifications may each yield internally consistent spectral types when compared to standards, these classifications probe physically distinct components of the stellar surface. In its ability to be mapped to a single effective temperature value, the concept of a unique spectral type thus loses coherence in the context of heavily spotted stars, especially when spots contribute a substantial fraction, up to $80\%$ in some of our models, of the total bolometric luminosity. 

In light of these results, we advocate for a paradigm shift in the treatment of PMS stars in both observational and theoretical contexts. Effective temperatures should not be inferred from spectral types alone unless the method explicitly models or corrects for spot contamination. Stellar evolutionary models must also be re-calibrated to account for the altered Effective Temperature–Luminosity–Radius relationships in spotted stars, as has begun to be explored in the context of magnetic and spotted models \citep[][]{Feiden2016, Somers2020}. 

\subsection{Recommendations for Inferring Fundamental Stellar Parameters of Spotted PMS Stars}

The results presented in this work strongly motivate a shift toward two-temperature spectral modeling for young, magnetically active stars. However, we recognize that many studies of PMS populations rely on more limited datasets (e.g., photometry, low-resolution spectroscopy) or aim to characterize large samples where full spectral modeling is impractical. Here we offer a set of practical recommendations for inferring stellar parameters in such contexts, with the goal of mitigating the most significant biases introduced by starspots.

For studies focused on mass and age estimation, our results suggest that using magnetic stellar evolution models, such as those developed by F16, without further corrections offers a substantial improvement over traditional non-magnetic tracks when working with uncorrected (i.e., spot-contaminated) effective temperatures. These models implicitly encode the effects of interior magnetic inhibition of convection, which mimics the cooling impact of surface inhomogeneities. As shown in Section \ref{sec:hr_comp}, this reduces discrepancies between spot-corrected and non-corrected parameters by 50–70\% in much of the $\rm 0.3\ M_{\odot} \leq M\leq 0.8\ M_{\odot}$ regime. Accordingly, we recommend that in the absence of spot corrections, the F16 tracks serve as the default choice for mass and age inference in active PMS stars.

However, this approach is not sufficient in all contexts. In particular, estimates of bolometric luminosity, radius, and surface gravity are far more sensitive to the shape of the spectral energy distribution than to effective temperature alone. Even when spot-corrected $\rm T_{eff}$ values are available, bolometric corrections based on single-temperature assumptions systematically misestimate the total luminosity, especially in stars where the cool component dominates the infrared output. As shown in Figure~\ref{fig:trends}, spot contributions can reach 70\% of the total bolometric flux in the 3350–4000~K range. Thus, any parameter that depends on spectral morphology, rather than $\rm T_{eff}$ alone, requires explicit modeling of the composite atmosphere.

We therefore caution against the use of empirical spectral type-$\rm T_{eff}$ conversions and bolometric corrections in spotted stars without first assessing the likelihood of significant spot coverage. Observational flags such as strong optical/NIR log g discrepancies, red-optical molecular band strengths, and apparent $\rm T_{eff}$ mismatches across wavelength regimes can serve as indirect indicators of spot contamination. When such indicators are present, single-temperature inferences should be treated with a healthy bit of skepticism. While two-temperature spectral modeling remains the most robust approach, use of magnetic evolutionary models like those of F16 can recover reasonably accurate masses and ages from spot-contaminated temperatures provided that care is taken to avoid applying these models in contexts where luminosity, radius, or SED shape are the observables used to infer other stellar parameters. We recommend that future observational pipelines integrate spot-aware modeling whenever possible and treat traditional parameter inference methods as conditionally valid approximations, not general solutions.

\subsection{Implications for Veiling and Accretion Diagnostics in Classical T Tauri Stars}

The systematic biases introduced by starspots in the spectra of WTTS have direct implications for the study of CTTS, where accurate characterization of the stellar photosphere is critical for isolating accretion emission. A common approach to estimating veiling in CTTS involves subtracting an empirical WTTS template to recover the excess continuum and infer the accretion luminosity. This method implicitly assumes that WTTS templates represent unspotted or minimally spotted stellar atmospheres. Our results suggest that many WTTS have significant spot coverage making this a poor assumption for the majority of would-be spectral templates.

As demonstrated in Section \ref{sec:spotfits}, the WTTS spectra themselves often reflect composite photospheres with spot filling factors ranging from 20\% to over 70\%, depending on stellar mass and evolutionary state. These spots not only alter the spectral morphology across optical and near-infrared wavelengths, but also significantly distort the spectral energy distribution. When such spotted spectra are used as templates for stellar contribution subtraction, the inferred excess continuum, and hence the accretion luminosity, can be systematically biased. This effect was examined in \citet{PérezPaolino2025}, where use of unspotted templates led to accretion luminosities being underestimated or overestimated by factors of a few to ten, depending on the mismatch between the template and the true composite spectrum of the CTTS. The discrepancy will be particularly acute in the 3350–4000~K regime, where spot contributions dominate the infrared flux and substantially dilute the photospheric continuum. Models for the underlying stellar emission in accreting systems similar to those presented here might provide a more physically realistic basis for veiling measurements and correction, particularly when the goal is to reconstruct the accretion shock emission or compute $\rm \dot{M}$ from UV or near-infrared excesses.

Moreover, bolometric corrections applied to the veiling continuum must be revisited in light of these results. Since the shape of the photospheric SED is altered by spots, assumptions about the excess-to-bolometric scaling based on single-temperature continua are no longer valid. This introduces further uncertainty into accretion rate estimates. For example, recent results by \citet{Pittman2022} suggest the presence of low energy density accretion columns covering large portions of the stellar surface, and contributing over 70\% of the accretion luminosity. Since these stars lie in the mass range of enhanced starspot activity, where starspot luminosity can reach 70\% of the photospheric luminosity, care should be taken not to confuse unsubtracted starspot contributions to the net stellar spectrum with a broad, warm blackbody excess from accretion. Taken together, these findings suggest that starspot contamination may be a hidden source of scatter, and even systematic error, in CTTS accretion diagnostics across the literature. As a result, the application of empirical templates must take these effects of starspots into consideration to remove or minimize their impact on measurements of mass accretion rates.

\section{Conclusion}
In this study, we have demonstrated that the traditional treatment of PMS stars as single-temperature objects leads to systematic biases in the inference of their effective temperatures, surface gravities, luminosities, masses, and ages. Using high-resolution, broad-band X-Shooter spectra of 56 WTTS, we constructed two-temperature models that explicitly account for surface inhomogeneities caused by cool starspots. These composite models outperform conventional single-temperature fits across optical and near-infrared wavelengths. Our results show that spot filling factors exceeding 50\% are common, with the spot component often dominating the flux at red-optical and infrared wavelengths. By recovering both photospheric and spot temperatures, as well as their respective surface gravities and filling factors, we offer a more physically consistent interpretation of WTTS spectra. These findings suggest that traditional spectral typing methods, whether based on indices or template matching, are insufficient to characterize spotted PMS stars.

A key outcome of this analysis is the demonstration that spot-corrected parameters shift stellar positions in the HR Diagram, often substantially. Single-temperature fits can overestimate effective temperatures by up to 700~K and underestimate surface gravities by 1–2~dex. This leads to significant biases in derived stellar masses and ages when using standard evolutionary models. Our results show that spot corrections raise inferred stellar masses by up to 80\% and increase ages by 0.3–0.5~dex, with the largest discrepancies occurring for stars in the 0.3–0.8 $M_{\odot}$ range (corresponding to 3350-4000~K). We attribute this to both the high spot filling factors found in this mass regime. These stars are at or near the transition to radiative core development, making them structurally and magnetically distinct from both lower and higher mass WTTSs.

While magnetic stellar evolution models, such as those by \citet{Feiden2016}, reduce some of the biases induced by starspots, we argue that they do not eliminate the need for explicit spot treatment. Magnetic suppression of convection in the stellar interior can mimic some effects of spot coverage on the HR~diagram, but cannot resolve the full range of spectral inconsistencies observed across wavelengths. Our results suggest that spot-corrected temperatures, being offset by as much as 500~K from those derived using unspotted models, {with an average difference of 200~K,} yield better internal consistency and provide more accurate placement in evolutionary model grids. These shifts help alleviate tensions when masses and ages are inferred from surface gravity and temperature diagnostics, shifting masses by as much as $40\%$ and ages by a factor of two. Moreover, by treating starspots as surface manifestations of large-scale magnetic topology, we identify a possible empirical link between magnetic geometry and stellar structure. Specifically, the transition from high spot coverage and axisymmetric dipole fields to lower spot coverage and complex, small-scale magnetic fields appears to coincide with the development of radiative cores, potentially demarcating a shift from convective to interface dynamo regimes in young stars. These results seem to confirm the predictions of \citet{Gregory2012}.

Beyond refining parameter estimation, our analysis has broader implications for studies of young stellar populations. Systematic biases in mass and age estimates propagate into the inferred IMF, star formation histories, and cluster age spreads. Our results indicate that a significant fraction (0.3 dex) of the apparent age spread in young star-forming regions may be attributable to spot-induced temperature biases rather than intrinsic age dispersion. While we caution against over-interpreting scatter in a heterogeneous sample drawn from multiple star-forming regions of different ages, the internal consistency achieved through spot-aware modeling suggests that these methods can reduce artificial age-mass trends commonly seen in HR diagram analyses. 

In future work, we aim to apply this framework to WTTSs with independently measured dynamical masses to test the fidelity of spot-corrected evolutionary tracks. Additionally, we plan to extend this approach to CTTSs, where veiling and accretion-driven continuum excess further complicate the spectral decomposition. A complementary test of the spot-corrected parameters presented here would be to compare model-inferred ages to those derived from lithium depletion. As lithium abundance traces interior temperatures rather than surface properties, such a comparison could provide an independent check on whether spot corrections yield more physically consistent ages. A detailed analysis of lithium equivalent widths and lithium depletion ages for this sample is relegated to future work. Ultimately, a physically motivated, multi-component treatment of stellar atmospheres that includes starspots, magnetic fields, and accretion signatures is necessary to build a self-consistent picture of pre-main sequence stellar evolution. The models and methodology presented here constitute a foundational step toward that goal. 

\begin{acknowledgements}
{Based on observations collected at the European Southern Observatory under ESO programmes 085.C-0764(A), 093.C-0506(A), 106.20Z8.004, 106.20Z8.006, 106.20Z8.008, 109.23D4.001, 110.23P2.001.}
\end{acknowledgements}

\software{{Astropy \citep{AstropyI, AstropyII, AstropyIII}}, Spextool \citep{cushing2004}, matplotlib \citep{Hunter2007}, Scipy 
\citep{jones2014}, Numpy \citep{VanderWalt2011}, lmfit \citep{Newville2014}, emcee \citep{Foreman-Mackey2013}, SpectRes \citep{2017arXiv170505165C}}

\restartappendixnumbering
\appendix
\section{Detectability Limits for Spectroscopic Fitting of Spotted Stars}
In this section we perform synthetic tests to determine the limits of our method. To do this, we first artificially degrade XSHOOTER spectra to lower Signal-to-Noise (SNR) and determine a noise floor for detectability of starspot signatures. Then we create synthetic spotted stars with increasingly close spot and photosphere temperatures to test whether our models can reproduce the input parameters.

For our first test, we selected the spectrum of RXJ1543.1-3920, given its well-behaved fits as demonstrated in Section \ref{sec:spotfits}. SNRs reported by \citet{Claes2024} across the relevant spectral range exhibit an unexpectedly broad variation for this source, ranging from 16 to 205 over the optical to near-infrared wavelengths. However, this spread appears inconsistent with our own inspection of the data, which indicates a minimum SNR of approximately 50. As a result, we adopt a conservative estimate for the true SNR of RXJ1543.1-3920, assuming a representative value of around 100. To degrade the spectrum noise was drawn from a normal distribution centered on the reported flux value, with a standard deviation defined as the flux divided by the desired SNR. We acknowledge that this approach neglects the intrinsic wavelength dependence of the SNR, which arises from variations in detector sensitivity, and instead assumes uniform efficiency across the X-Shooter spectral range. Nevertheless, we consider this approximation sufficient to capture the essential limitations of our method. We artificially degraded the RXJ1543.1-3920 spectrum to target SNRs of 50, 30, 20, 10, 5, and 2 by adding random noise to each spectral point and re-fit this modified spectrum. 

\begin{table*}[ht]
\caption{Detectability Limits of Two-Temperature Fits for Hotter Stars.}\label{tab:degrade}
\begin{center}
\begin{tabular}{lccccc}
\hline
\multicolumn{6}{c}{Varying SNR fit to RXJ1543.1-3920} \\
\colhead{SNR} & \colhead{$\rm f_{ spot}$} & \colhead{$\rm T_{phot}$} & \colhead{$\rm T_{spot}$} & \colhead{$\rm \log g$ (phot)} & \colhead{$\rm \log g$ (spot)}\\
\colhead{} & \colhead{} & \colhead{(K)} & \colhead{(K)} \\
\hline
100 & $0.69\pm{0.02}$ & $4634\pm{40}$ & $3714\pm{40}$ & $4.41\pm{0.09}$ & $5.00\pm{0.10}$\\
50 & $0.75\pm{0.05}$ & $4748\pm{70}$ & $3749\pm{55}$ & $4.67\pm{0.17}$ & $4.91\pm{0.10}$\\
30 & $0.74\pm{0.06}$ & $4753\pm{75}$ & $3733\pm{87}$ & $4.74\pm{0.16}$ & $4.94\pm{0.17}$\\
20 & $0.70\pm{0.04}$ & $4654\pm{75}$ & $3732\pm{45}$ & $4.42\pm{0.17}$ & $5.06\pm{0.09}$\\
10 & $0.78\pm{0.04}$ & $4723\pm{85}$ & $3894\pm{45}$ & $4.17\pm{0.25}$ & $4.82\pm{0.13}$\\
5 & $0.71\pm{0.02}$ & $4637\pm{140}$ & $3554\pm{66}$ & $4.39\pm{0.30}$ & $5.07\pm{0.07}$\\
2 & $0.45\pm{0.29}$ & $4627\pm{45}$ & $2000\pm{50}$ & $3.50\pm{0.58}$ & $4.26\pm{0.34}$\\
\hline
\multicolumn{6}{c}{Fixed SNR = 30 varying $\rm T_{phot}$ and $\rm T_{spot}$} \\
\colhead{$\rm T_{phot}, T_{spot}$} & \colhead{$\rm f_{ spot}$} & \colhead{$\rm T_{phot}$} & \colhead{$\rm T_{spot}$} & \colhead{$\rm \log g$ (phot)} & \colhead{$\rm \log g$ (spot)}\\
\colhead{(K)} & \colhead{} & \colhead{(K)} & \colhead{(K)} \\
\hline
4000, 3000 & $0.65\pm{0.20}$ & $3998\pm{24}$ & $2970\pm{32}$ & $3.61\pm{0.05}$ & $4.45\pm{0.07}$ \\
3900, 3100 & $0.66\pm{0.03}$ & $3910\pm{27}$ & $3099\pm{34}$ & $3.63\pm{0.04}$ & $4.54\pm{0.07}$ \\
3800, 3200 & $0.68\pm{0.06}$ & $3815\pm{38}$ & $3233\pm{43}$ & $3.52\pm{0.08}$ & $4.52\pm{0.06}$ \\
3700, 3300 & $0.61\pm{0.08}$ & $3683\pm{37}$ & $3278\pm{36}$ & $3.53\pm{0.13}$ & $4.50\pm{0.07}$ \\
3600, 3400 & $0.70\pm{0.09}$ & $3613\pm{47}$ & $3412\pm{25}$ & $3.51\pm{0.22}$ & $4.48\pm{0.07}$ \\
3500, 3500 & $0.75\pm{0.09}$ & $3604\pm{35}$ & $3437\pm{21}$ & $3.46\pm{0.22}$ & $4.42\pm{0.07}$ \\
\hline 
\multicolumn{6}{c}{Fixed SNR = 100 varying $\rm T_{phot}$ and $\rm T_{spot}$} \\
\colhead{$\rm T_{phot}, T_{spot}$} & \colhead{$\rm f_{ spot}$} & \colhead{$\rm T_{phot}$} & \colhead{$\rm T_{spot}$} & \colhead{$\rm \log g$ (phot)} & \colhead{$\rm \log g$ (spot)}\\
\colhead{(K)} & \colhead{} & \colhead{(K)} & \colhead{(K)} \\
\hline
4000, 3000 & $0.65\pm{0.01}$ & $4002\pm{15}$ & $3004\pm{18}$ & $3.63\pm{0.03}$ & $4.47\pm{0.04}$ \\
3900, 3100 & $0.65\pm{0.02}$ & $3908\pm{15}$ & $3096\pm{20}$ & $3.61\pm{0.03}$ & $4.45\pm{0.04}$ \\
4000, 3000 & $0.65\pm{0.02}$ & $3908\pm{15}$ & $3096\pm{20}$ & $3.61\pm{0.03}$ & $4.45\pm{0.04}$ \\
3800, 3200 & $0.70\pm{0.02}$ & $3821\pm{15}$ & $3244\pm{19}$ & $3.55\pm{0.05}$ & $4.48\pm{0.04}$ \\
3700, 3300 & $0.64\pm{0.06}$ & $3693\pm{27}$ & $3294\pm{29}$ & $3.65\pm{0.10}$ & $4.51\pm{0.05}$ \\
3600, 3400 & $0.72\pm{0.12}$ & $3603\pm{50}$ & $3416\pm{22}$ & $3.48\pm{0.21}$ & $4.45\pm{0.08}$ \\
3500, 3500 & $0.67\pm{0.10}$ & $3609\pm{42}$ & $3413\pm{22}$ & $3.52\pm{0.23}$ & $4.49\pm{0.07}$ \\
\hline
\end{tabular}
\end{center}
\end{table*}

For spectra degraded to SNR of 10, the best-fit parameters remain consistent with those derived from the high-SNR data, with deviations lying within the 1-$\sigma$ posterior uncertainties (Table~\ref{tab:degrade}. However, at SNR~=~5, the fits begin to diverge from the true solution, and by SNR~=~2, the retrieved parameters become unreliable. In this case, the model converges on a spot temperature of 2000~K, a value that is physically implausible and clearly unconstrained by the data, while the spot filling factor and surface gravities change drastically from the higher-SNR fits. The full results of this experiment are shown in Table~\ref{tab:degrade}.

\begin{figure}[h]
\centering
\includegraphics[width=1\linewidth]{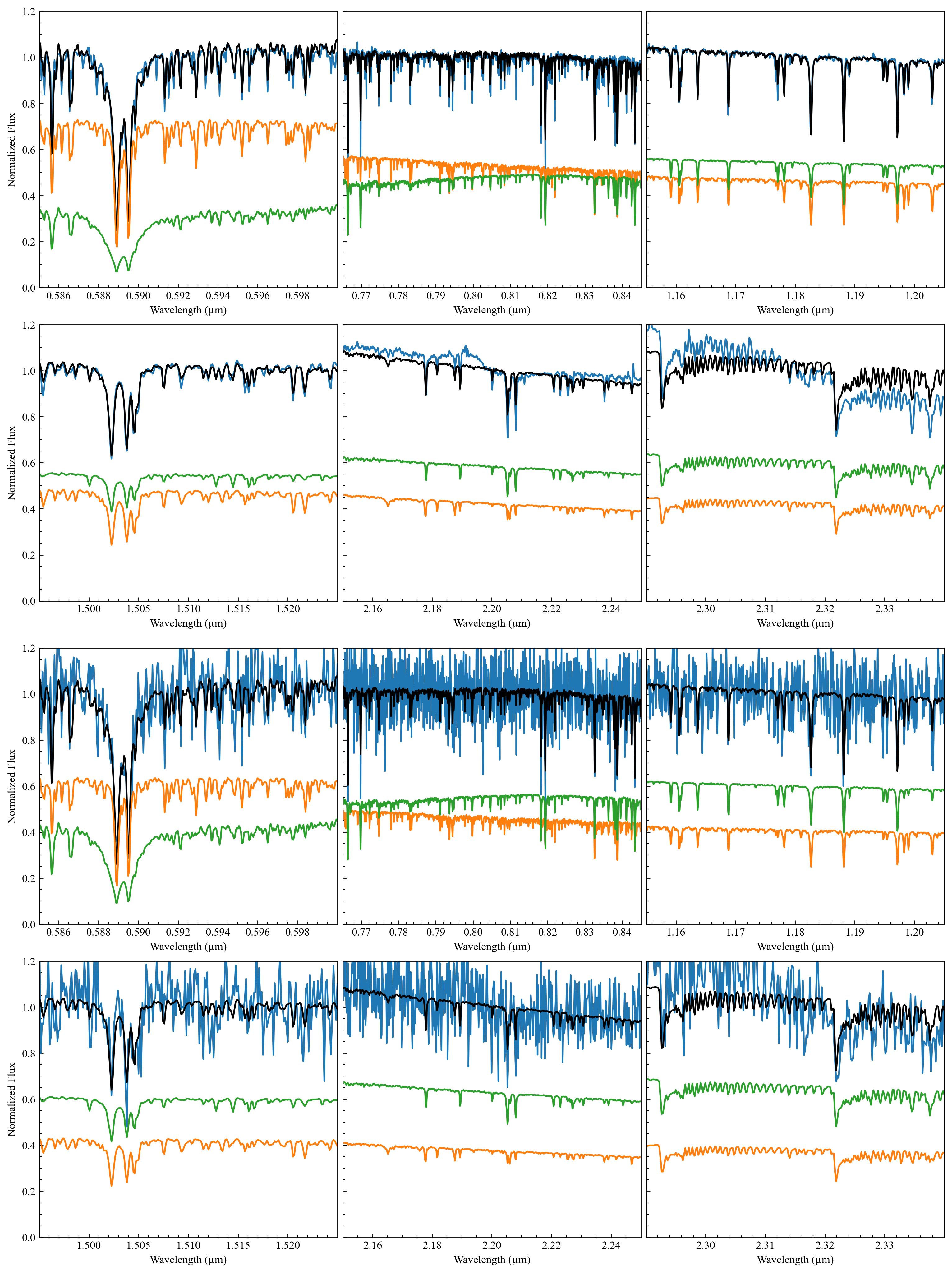}
\caption{\textbf{Fitting Degraded RXJ1543.1-3920 Spectra}. Plotted are two fits to the same RJX1543.1-3920 spectrum at different signal to noise ratios. Top: SNR$\approx$100, bottom: SNR = 10. Colors are the same as in Figure~\ref{fig:twofits}.} \label{fig:limits}
\end{figure}

We then performed a second experiment where models of spotted stars were constructed with a fixed $\rm f_{spot}=0.65$, $\rm \log g\ (phot)=3.6$, $\rm \log g\ (spot)=4.5$, $\rm v\ sini=30\ km/s$, and $\rm v_r=30\ km/s$, but with decreasing temperature contrast from $\rm T_{phot}=4000$, $\rm T_{spot}=3000$ to $\rm T_{phot}=3500$, $\rm T_{spot}=3500$ in steps of 100 K.

As the temperature contrast narrows, the stellar spectrum becomes increasingly dominated by continuum shape rather than discrete spectral features. In this regime, the two-component model effectively becomes degenerate with a single-temperature photosphere, and the spot signature is no longer recoverable even at high SNR. This defines a physical detectability threshold set not by data quality, but by atmospheric homogeneity, at the X-Shooter resolution. At ($\rm T_{phot}$, $\rm T_{spot}$) = (3500, 3500), the model formally recovers two distinct components, but their separation is within observational uncertainty, and the fit is statistically indistinguishable from a single-temperature spectrum. We conclude that the closest temperature difference we can measure between the starpots and the photosphere even at high filling factors of $\rm f_{spot}=0.65$ is on the order of $\approx 150$~K. Spot temperatures within $\approx 150$~K of the photospheric temperature cannot be robustly recovered, even with high SNR and broad spectral coverage.

\begin{table*}[ht]
\caption{Detectability Limits of Two-Temperature Fits for Colder Stars.}\label{tab:degrade2}
\begin{center}
\begin{tabular}{lccccc}
\hline
\multicolumn{6}{c}{Varying SNR fit to TWA~7} \\
\colhead{SNR} & \colhead{$\rm f_{ spot}$} & \colhead{$\rm T_{phot}$} & \colhead{$\rm T_{spot}$} & \colhead{$\rm \log g$ (phot)} & \colhead{$\rm \log g$ (spot)}\\
\colhead{} & \colhead{} & \colhead{(K)} & \colhead{(K)} \\
\hline
100 & $0.31\pm{0.02}$ & $3399\pm{10}$ & $2786\pm{40}$ & $3.84\pm{0.02}$ & $5.26\pm{0.09}$\\
50 & $0.30\pm{0.02}$ & $3398\pm{15}$ & $2771\pm{35}$ & $3.85\pm{0.03}$ & $5.29\pm{0.10}$\\
30 & $0.30\pm{0.02}$ & $3399\pm{15}$ & $2786\pm{37}$ & $3.84\pm{0.03}$ & $5.29\pm{0.11}$\\
20 & $0.29\pm{0.02}$ & $3399\pm{20}$ & $2762\pm{38}$ & $3.88\pm{0.05}$ & $5.42\pm{0.07}$\\
10 & $0.32\pm{0.02}$ & $3399\pm{20}$ & $2654\pm{28}$ & $3.89\pm{0.02}$ & $5.39\pm{0.06}$\\
5 & $0.42\pm{0.02}$ & $3398\pm{20}$ & $2495\pm{28}$ & $3.86\pm{0.09}$ & $4.08\pm{0.07}$\\
2 & $0.74\pm{0.05}$ & $3608\pm{20}$ & $2430\pm{23}$ & $3.43\pm{0.04}$ & $4.12\pm{0.05}$\\
\hline
\multicolumn{6}{c}{Fixed SNR = 30 varying $\rm T_{phot}$ and $\rm T_{spot}$} \\
\colhead{$\rm T_{phot}, T_{spot}$} & \colhead{$\rm f_{ spot}$} & \colhead{$\rm T_{phot}$} & \colhead{$\rm T_{spot}$} & \colhead{$\rm \log g$ (phot)} & \colhead{$\rm \log g$ (spot)}\\
\colhead{(K)} & \colhead{} & \colhead{(K)} & \colhead{(K)} \\
\hline
3100, 3100 & $0.10\pm{0.06}$ & $3108\pm{10}$ & $2993\pm{70}$ & $3.79\pm{0.05}$ & $4.59\pm{0.21}$ \\
3200, 3000 & $0.31\pm{0.04}$ & $3205\pm{20}$ & $2999\pm{14}$ & $3.61\pm{0.03}$ & $4.51\pm{0.07}$ \\
3300, 2900 & $0.30\pm{0.05}$ & $3292\pm{22}$ & $2933\pm{23}$ & $3.59\pm{0.03}$ & $4.55\pm{0.14}$ \\
3400, 2800 & $0.33\pm{0.02}$ & $3405\pm{15}$ & $2756\pm{33}$ & $3.60\pm{0.02}$ & $4.39\pm{0.07}$ \\
\hline 
\multicolumn{6}{c}{Fixed SNR = 100 varying $\rm T_{phot}$ and $\rm T_{spot}$} \\
\colhead{$\rm T_{phot}, T_{spot}$} & \colhead{$\rm f_{ spot}$} & \colhead{$\rm T_{phot}$} & \colhead{$\rm T_{spot}$} & \colhead{$\rm \log g$ (phot)} & \colhead{$\rm \log g$ (spot)}\\
\colhead{(K)} & \colhead{} & \colhead{(K)} & \colhead{(K)} \\
\hline
3100, 3100 & $0.32\pm{0.11}$ & $3104\pm{10}$ & $3095\pm{11}$ & $3.59\pm{0.09}$ & $4.46\pm{0.014}$ \\
3200, 3000 & $0.33\pm{0.04}$ & $3211\pm{20}$ & $2989\pm{21}$ & $3.58\pm{0.04}$ & $4.42\pm{0.07}$ \\
3300, 2900 & $0.30\pm{0.06}$ & $3298\pm{23}$ & $2921\pm{32}$ & $3.59\pm{0.05}$ & $4.50\pm{0.12}$ \\
3400, 2800 & $0.29\pm{0.03}$ & $3396\pm{15}$ & $2810\pm{32}$ & $3.60\pm{0.02}$ & $4.53\pm{0.09}$ \\
\hline
\end{tabular}
\end{center}
\end{table*}

{We repeated these experiments for stars of colder temperatures closer to the fully convective boundary to ensure that our observed trends are not an artifact of limitations in the fitting routine. The result of these is shown in Table~\ref{tab:degrade2}. We find similar sensitives for these lower temperature stars, concluding that the trends we find across temperature are real and not affected by systematics.
}
\pagebreak
\bibliography{spots_templates.bib}{}
\bibliographystyle{aasjournal}

\end{document}